%% file: paper.tex
\newif \ifdraft \drafttrue 
\documentclass[10pt,twocolumn,sigconf]{acmart}

\newif \ifappendix \appendixfalse

\usepackage{booktabs} % For formal tables

\author{Vibhaalakshmi Sivaraman}
\affiliation{
	\institution{Princeton University}
}

\author{Srinivas Narayana}
\affiliation{
	\institution{MIT CSAIL}
}

\author{Ori Rottenstreich}
\affiliation{
	\institution{Princeton University}
}

\author{S. Muthukrishnan}
\affiliation{
	\institution{Rutgers University}
}

\author{Jennifer Rexford}
\affiliation{
	\institution{Princeton University}
}

\renewcommand{\shortauthors}{V. Sivaraman et al.}

\copyrightyear{2017}
\acmConference[SOSR '17]{ACM Symposium on SDN Research}{April 3-4, 2017}{Santa Clara, CA} 
\acmISBN{978-1-4503-4947-5/17/04}
\acmPrice{\$15.00}
\acmDOI{http://dx.doi.org/10.1145/3050220.3050239}

\usepackage[normalem]{ulem}
\usepackage{amsmath, amssymb}
\usepackage{fullpage}
\usepackage{framed}
\usepackage{graphicx}
\usepackage{wrapfig}
\usepackage[linesnumbered,ruled]{algorithm2e}
\usepackage{caption}
\usepackage{wrapfig}
\usepackage{subcaption}
\usepackage{lipsum}
\usepackage[titletoc]{appendix}
\usepackage{hyperref} % http://ctan.org/pkg/hyperref
\hypersetup{%
  colorlinks = true,
  linkcolor  = black
}
\usepackage{csquotes}
\usepackage{xspace}
\usepackage{ragged2e}
\usepackage{fixltx2e}
\usepackage{amsmath,amsthm,amsfonts}
\usepackage{mdframed}
\usepackage{stmaryrd}
\usepackage{calc}
\usepackage[makeroom]{cancel}
\usepackage{algpseudocode}
\ifappendix \usepackage[ruled,vlined]{algorithm2e} \fi
\usepackage{makecell}
\usepackage{lineno}

\usepackage{enumitem}
\usepackage[justification=justified, font=small]{caption}
\usepackage[export]{adjustbox}
\usepackage{bm}
\usepackage{amstext}
\usepackage{tabularx}
\usepackage{balance}
\usepackage{authblk}
\usepackage[export]{adjustbox}

\usepackage{listings}
\usepackage{color}

\definecolor{codegreen}{rgb}{0,0.6,0}
\definecolor{codegray}{rgb}{0.5,0.5,0.5}
\definecolor{codepurple}{rgb}{0.58,0,0.82}
\definecolor{backcolour}{rgb}{0.95,0.95,0.92}

\graphicspath{ {figures/} }

\makeatletter
\renewcommand\AB@affilsepx{ , \protect\Affilfont}
\makeatother

\setlist[enumerate,1]{%
  label=\arabic*.,
}

\newlist{inlinelist}{enumerate*}{1}
\setlist*[inlinelist,1]{%
  label=(\roman*),
}

\newcommand{\ignore}[1]{}

\newcommand{\eg}{{\em e.g., }}
\newcommand{\etal}{{\em et al.}}
\newcommand{\vs}{{\em vs. }}
\newcommand{\ie}{{\em i.e., }}
\newcommand{\Sec}[1]{\S\ref{#1}}

\newcommand{\TheSystem}{HashPipe\xspace}
\newcommand{\Baseline}{HashParallel\xspace}
\newcommand{\spacesaving}{space saving\xspace}
\newcommand{\SpaceSaving}{Space Saving\xspace}
\newcommand{\Spacesaving}{Space saving\xspace}
\newcommand{\NewPara}[1]{\noindent{\bf #1}}
\newcommand{\Alg}[1]{Algorithm~\ref{algo:#1}}
\newcommand{\Fig}[1]{Fig.~\ref{fig:#1}}

\usepackage[hang,flushmargin]{footmisc} 
\usepackage{textcomp}
\newcommand{\cref}[1]{\textsection\ref{#1}}
\newsavebox\CBox
\newcommand\hcancel[2][0.5pt]{%
	\ifmmode\sbox\CBox{$#2$}\else\sbox\CBox{#2}\fi%
	\makebox[0pt][l]{\usebox\CBox}%  
	\rule[0.5\ht\CBox-#1/2]{\wd\CBox}{#1}}
\begin{document}
%%\title{Smoking Out the Heavy-Hitter Flows with HashPipe}
\title{Heavy-Hitter Detection Entirely in the Data Plane}
\thispagestyle{empty}

%\authornote{Dr.~Trovato insisted his name be first.}
%\affiliation{%
%  \institution{Institute for Clarity in Documentation}
%  \streetaddress{P.O. Box 1212}

%\author[1]{Vibhaalakshmi Sivaraman}
%\author[2]{Srinivas Narayana}
%\author[1]{Ori Rottenstreich}
%\author[3]{\\S. Muthukrishnan}
%\author[1]{Jennifer Rexford}
%\affil[1]{Princeton University}
%\affil[2]{MIT CSAIL}
%\affil[3]{Rutgers University}
%\setcounter{Maxaffil}{0}
%\renewcommand\Affilfont{\small}

\input{abstract}

% \begin{CCSXML}e
% <concept_id>10003033.10003068</concept_id>
% <concept_desc>Networks~Network algorithms</concept_desc>
% <concept_id>10003033.10003099.10003105</concept_id>
% <concept_desc>Networks~Network monitoring</concept_desc>
% <concept_id>10003033.10003099.10003102</concept_id>
% <concept_desc>Networks~Programmable networks</concept_desc>
% <concept_significance>300</concept_significance>

%ccsdesc[500]{Networks~Network algorithms}
%ccsdesc[500]{Networks~Network monitoring}
%ccsdesc[300]{Networks~Programmable networks}

 %e filled

\maketitle

%\AcmCopyright
%\ToAppear
%\begin{sloppypar}

%\ifthenelse{\equal{\onlyAbstract}{no}}{% !onlyAbstract

\input{introduction}

\input{problem}

\input{related}
\input{algorithm}

\input{prototype}
\input{evaluation}

\input{other-related}
\input{conclusion}
\label{lastpage}

%\begin{appendices}
%\input{analysis}
%\end{appendices}

%\end{sloppypar}

%\vspace{-0.1in}
%\section*{Acknowledgments}
% Comments for people we need to acknowledge in the final version.
\begin{acks}
  We thank the anonymous SOSR reviewers, Anirudh Sivaraman, and Bharath
  Balasubramanian for their feedback.  This
  work was supported partly by NSF grant CCF-1535948, DARPA grant
  HR0011-15-2-0047 and gifts from
  Intel and Huawei. We also thank the industrial members of the MIT Center for
  Wireless Networks and Mobile Computing (Wireless@MIT) for their support.

\end{acks}

%\pagebreak

\clearpage
\setlength{\parskip}{-1pt}
\setlength{\itemsep}{-1pt}
 \footnotesize % SPACE

\bibliographystyle{abbrv}
\begin{small}
\bibliography{paper}
\end{small}
%\bibliographystyle{abbrvnat_noaddr} % SPACE
%\theendnotes % ENDNOTES

%}
{% onlyAbstract
}

\ifappendix
\clearpage
\appendix
\input{appendix}
\fi
\end{document}

%% file: abstract.tex
\begin{abstract}

% doing this entirely in the dataplane
Identifying the ``heavy hitter'' flows or flows with large traffic volumes in
the data plane is important for several applications \emph{e.g.,} flow-size
aware routing, DoS detection, and traffic engineering.
%Understanding which  traffic flows consume the most bandwidth is useful for load
%balancing, detecting DoS attacks, and traffic engineering. Identifying these
%``heavy hitter'' flows in the data plane enables applications that distinguish
%between packets belonging to heavy and light flows at packet-processing time,
%\emph{e.g.,}
%to route heavy flows or DoS traffic differently. 
However, measurement in the data plane is constrained by the need for line-rate processing (at 10-100Gb/s) and
limited memory in switching hardware. We propose HashPipe, a heavy hitter detection algorithm using emerging programmable data planes.
HashPipe implements a pipeline of hash tables which retain counters for heavy
flows in the tables while evicting lighter flows over time.
We prototype HashPipe in P4 and
evaluate it with packet traces from an ISP backbone link and a data center. On
the ISP trace, we
find that
HashPipe identifies 95\% of the 300 heaviest flows with less than 80KB of memory
on a trace that contains 400,000 flows.
\end{abstract}

%% file: introduction.tex
\section{Introduction}
\label{sec:intro}

%Accurate measurement of network traffic is important for management tasks,
%%  such as traffic engineering, provisioning capacity, and diagnosing and
%% stopping DoS attacks.
%as lack of visibility into the network can lead to poor performance and costly
%outages.
%% Further, as packet-processing becomes increasingly flexible
%% \cite{mckeown2008openflow, bosshart2014p4, domino}, it is important to drive
%% fine-grained network control through accurate monitoring data. 
%As a result, both industry and academia have actively invested in improving the
%state of the art~\cite{cisco-netflow, estan2002new, int, yu2013software,
 % li2016flowradar, univmon}.

Many network management applications can benefit from finding the set of flows
contributing significant amounts of traffic to a link: for example, to relieve
link congestion~\cite{microTE}, to plan network
capacity~\cite{att-deriving-traffic-demands}, to detect network anomalies and
attacks~\cite{network-wide-anomalies}, or to cache forwarding table
entries~\cite{LossyToN}.
Further, identifying such ``heavy hitters'' at small time scales (comparable to
traffic variations~\cite{hedera, microTE}) can enable dynamic routing of heavy
flows~\cite{devoflow, planck} and dynamic flow scheduling~\cite{pifo}.

It is desirable to run heavy-hitter monitoring at all switches in the network
all the time, to respond quickly to short-term traffic variations. {\em Can
  packets belonging to heavy flows be identified as the packets are processed in
  the switch}, so that switches may treat them specially?

Existing approaches to monitoring heavy items make it hard to achieve
reasonable accuracy at acceptable overheads (\Sec{sec:related}). While packet
{\em sampling} in the form of NetFlow ~\cite{cisco-netflow} is widely deployed,
the CPU and bandwidth overheads of processing sampled packets in software make
it infeasible to sample at sufficiently high rates (sampling just 1 in 1000
packets is common in practice~\cite{sflow-sampling-rate}). An alternative
is to use sketches, \eg \cite{cormode2005improved, li2016flowradar,
  yu2013software, univmon} that {\em hash} and {\em count} all packets in switch
hardware. However, these systems incur a large memory overhead to retrieve the heavy hitters
%required to estimate counts for {\em all} flows-
--- ideally, we wish to use memory proportional to the number of the heavy flows (say the top hundred). There may be tens of thousands of
active flows any minute on an ISP backbone link (\Sec{sec:evaluation}) or a data
center top-of-rack switch \cite{theo-dc-traffic}.

Emerging programmable switches~\cite{RMT,barefoot-tofino,cavium-xpliant} allow
us to do more than sample, hash, and count, which suggests opportunities to run
novel algorithms on switches. While running at line rates of
10-100 Gbps per port over 10-100 ports, these switches can be programmed to
maintain state over multiple packets, \eg to keep flow identifiers as keys
along with counts. Stateful manipulations can also be {\em pipelined}
over multiple stages, with the results carried by the packet from one stage to
the next and compared to stored state in subsequent stages.

However, switches are also constrained by the need to maintain high
packet-processing throughput, having:
\begin{itemize}
\item a deterministic, small time budget (1 ns \cite{RMT}) to manipulate state
  and process packets at each stage;
\item a limited number of accesses to memory storing state at each stage (typically just one read-modify-write);
\item a limited amount of memory available per stage (\eg 1.4MB shared across forwarding and monitoring~\cite{RMT}); 
%\item a limited amount of computation and boolean operations that are doable at each stage; and
\item a need to move most packets just once through the pipeline, to
  avoid stalls and reduced throughput
  (``feed-forward''~\cite{minions}).
\end{itemize}
We present \TheSystem,  an algorithm to track the $k$ heaviest flows with high
accuracy (\Sec{sec:algorithm}) within the features and constraints of
programmable switches. \TheSystem maintains both the
flow identifiers (``keys'') and counts of heavy flows in the switch, in a
pipeline of hash tables. 
%Packets flow from one stage to next and at each stage, the ``state'' carried by the packet maybe modified by the counts in that hash stage. This is distinct for example from generic multistage algorithms where the information passed between stages is contained in the program variables, whereas here the information flow is via the packets.  
%We design a new heavy hitter detection algorithm using \TheSystem as follows. 
When a packet hashes to a location in the first stage of the pipeline,
its counter is updated (or initialized) if there is a {\em hit} (or an empty
slot). If there is a {\em miss}, the new key is inserted at the expense of the
existing key. In all downstream stages, the key and count of the
item just evicted are carried along with the packet. The carried key is
looked up in
the current stage's hash table. Between the key looked up in the hash table and
the one carried, the key with the {\em larger count} is retained in the hash
table, while the other is either carried along with the packet to the next
stage, or totally removed from the switch if the packet is at the last
stage. Hence, \TheSystem ``smokes out'' the heavy keys within the limited
available memory, using pipelined operation to sample multiple locations in the
hash tables, and evicting lighter keys in favor of heavier
keys, with updates to exactly one location per stage. %% Our algorithm is
%% inspired by
%% the strategy used by the {\em space-savings} algorithm
%% \cite{metwally2005efficient}.\jrex{mention of the space-savings algorithm
%% arises abruptly, leaving the reader wondering how similar it is.  maybe we
%% could mention it earlier as a transition from the point that we need to keep
%% space proportional to k.}

We prototype \TheSystem  in P4~\cite{bosshart2014p4} (\Sec{sec:prototype}) and
test it on the public-domain behavioral switch model \cite{p4-bm}. %% We also
%% provide some basic analytical bounds for the estimation errors of our algorithm
%% (\Sec{sec:analysis}). 
We evaluate \TheSystem with packet traces obtained from an ISP backbone link
(from CAIDA) and a data
center, together containing over 500 million packets.
We show that \TheSystem can provide high accuracy
(\Sec{sec:evaluation}).
In the backbone link trace, \TheSystem incurs less than 5\% false negatives and
0.001\% false positives when reporting 300 heavy hitters (keyed by transport
five-tuple) with just 4500 counters (less than 80KB) overall, while the trace
itself contains 400,000 flows.
The errors both in false negatives and flow count estimations are lower among
the heavier flows in the top $k$.
At 80KB of memory, \TheSystem outperforms
sample and hold~\cite{estan2002new} by over $15\%$ and an enhanced version of the count-min sketch~\cite{cormode2005improved}
by $3$-$4\%$ in terms of false negatives.

%% file: problem.tex
\section{Background on Heavy-Hitter Detection}
\subsection{Problem Formulation}
\label{sec:problem}

\NewPara{Heavy hitters.} ``Heavy hitters'' can refer to all flows that are
larger (in number of packets or bytes) than a fraction $t$ of the total packets
seen on the link (the ``threshold-$t$'' problem). Alternatively, the heavy
hitters can be the top $k$ flows by size (the ``top-$k$''
problem). Through the rest of this paper, we use the ``top-$k$'' definition. %The two problems are nearly interchangeable, where a reasonable
%threshold $t$ can be used to identify the $k$ largest flows, and vice versa.

\NewPara{Flow granularity.} Flows can be defined at various levels of
granularity, such as IP address (\ie host), transport port
number (i.e., application), or five-tuple (\ie transport connection). With a
finer-grained notion of flows, the size and number of keys grows, requiring more
bits to represent the key and more entries in the data structure to track 
the heavy flows accurately.

\NewPara{Accuracy.} A technique for detecting heavy hitters may have false
positives (\ie reporting a non-heavy flow as heavy), false negatives (\ie not
reporting a heavy flow), or an error in estimating the sizes of heavy flows.
The impact of errors depends on the application.  For example, if the switch
performs load balancing in the data plane, a few false negatives may be
acceptable, especially if those heavy flows can be detected in the next time
interval.  As another example, if the network treats heavy flows as suspected
denial-of-service attacks, false positives could lead to unnecessary alarms that
overwhelm network operators. When comparing approaches in our evaluations, we
consider all three metrics.
%% and penalize approaches with a very high rate of either false positives or
%% false negatives.

\NewPara{Overhead.} The overhead on the switch includes the total amount of
memory for the data structure, the number of matching stages used in the switch
pipeline (constrained by a switch-specific maximum, say 16~\cite{RMT}). The
algorithms are constrained by the nature and amount of memory access and
computation per match stage (\eg computing hash functions). On high-speed links,
the number of active five-tuple flows per minute can easily be in the tens of
thousands, if not more. Our central goal is to maintain data-plane state that is proportional to the target number $k$ of heavy hitters (\eg $5k$ or
$10k$), rather than the number of active flows. In addition, we would like
to use as few pipeline stages as possible, since the switch also needs to
support other functionality related to packet forwarding and access control.

%% file: related.tex
\subsection{Existing Solutions}\label{sec:related}

The problem of finding heavy flows in a network is an instance of the ``frequent
items'' problem, which is extremely well studied in the streaming algorithms
literature~\cite{hh-cormode-survey}.
While high accuracy and low overhead are essential to any heavy hitter detection
approach, we are also specifically interested in implementing these algorithms
within the constraints of emerging programmable switches
(\Sec{sec:intro}).
We classify the approaches into two broad categories, {\em sampling} and {\em
  streaming,} discussed below.
%% \jrex{I think it'd be better to start with sketches like count-min sketch that
%%   make no effort to thin out or evict small flows, so you can transition from
%%   that to schemes that do such as count-min-with-cache, sample-and-hold, and
%%   finally space-savings.}

\NewPara{Packet sampling} using NetFlow~\cite{cisco-netflow} and sFlow~\cite{sflow}
is commonly implemented in routers today. These technologies record a subset of
network packets by sampling, and send the sampled records to collectors for
analysis. To keep packet processing overheads and data collection bandwidth low,
NetFlow configurations in practice use aggressively low sampling probabilities,
\eg 1\% or even 0.01\% \cite{sflow-sampling-rate, netflow-sampling-rate}. Such
undersampling can impact the estimation accuracy.

Sample and hold \cite{estan2002new} enhances the accuracy of packet sampling by
keeping counters for {\em all} packets of a flow in a ``flow table,'' once a
packet from that flow is sampled. However, implementing the flow table is
challenging, since the flow table requires a memory that is %sufficiently fast (to be read/written to by each packet) 
sufficiently large (to hold the necessary flow
entries).

Designing a large flow table\footnote{Large exact-match lookups are typically
  built with SRAM, as large TCAMs are expensive.} for fast packet {\em lookup}
is well-understood: hash table implementations are already common in switches,
for example in router forwarding tables and NetFlow~\cite{li2016flowradar}.
However, it is challenging to handle hash collisions when {\em adding} flows to
the flow table, when a packet from a new flow is sampled.
Some custom hardware solutions combine the hash table with a ``stash'' that
contains the overflow, \ie colliding entries, from the hash
table~\cite{sram-tcam-learning}. But this introduces complexity in the switch
pipeline, and typically involves the control plane to add entries into
the stash memory.
Ignoring such complexities, we evaluate sample and hold in \Sec{sec:evaluation}
by liberally allowing the flow table to lookup packets {\em anywhere} in a {\em
  list} that can grow up to a fixed size.

\NewPara{Streaming algorithms} implement data structures with bounded memory size
and processing time per packet, while processing {\em every packet} in a large
stream of packets in one pass. The algorithms are designed with provable
accuracy-memory tradeoffs for specific statistics of interest over the
packets. While these features make the algorithms attractive for network
monitoring, the specific algorithmic operations on each packet determine
feasibility on switch hardware.

Sketching algorithms like count-min sketch~\cite{cormode2005improved} and count
sketch~\cite{charikar2002finding} use per-packet operations such as hashing on
packet headers, incrementing counters at hashed locations, and determining the
minimum or median among a small number of the counters that were hashed
to. These operations can all be efficiently implemented on switch
hardware~\cite{yu2013software}.
However, these algorithms do not track the flow identifiers of packets, and hash
collisions make it challenging to ``invert'' the sketch into the constituent
flows and counters.
Simple workarounds like collecting packets that have high flow count estimates
in the sketch could result in significant bandwidth overheads, since most
packets from heavy flows will have high estimates.

Techniques like group testing~\cite{group-testing}, reversible sketches
\cite{schweller2004reversible}, and FlowRadar~\cite{li2016flowradar} can decode
keys from hash-based sketches. 
However, it is challenging to read off an accurate counter value for a packet in
the switch pipeline itself since the decoding happens off the fast packet-processing
path.
%
%
%Further, all the above approaches use the available memory to estimate {\em all}
%flows---not just the heavy ones.
%%(i) take up too many counter updates per packet
%%(ii) use too many counters overall too! accuracy-memory tradeoff not great

Counter-based algorithms~\cite{misra1982finding,manku2002approximate} focus on measuring the heavy
items, maintaining a table of flows and corresponding counts. They employ
per-counter increment and subtraction operations, but potentially all counters
in the table are updated during some flow insertions.
Updating multiple counters in a single stage is challenging within the deterministic time budget for each packet.
%Such worst-case table updates are not possible within the deterministic time budget for each packet.

\Spacesaving~\cite{metwally2005efficient} is a counter-based algorithm
that only uses $O(k)$ counters to track $k$ heavy flows, achieving the best
memory usage possible for a deterministic heavy-hitter algorithm. \Spacesaving
only updates one counter per packet, but requires finding the item with the
minimum counter value in the table. Unfortunately, scanning the entire table on
each packet, or finding the minimum in a table efficiently, is not directly
supported on emerging programmable hardware (\Sec{sec:intro}). Further,
maintaining data structures like sorted linked
lists~\cite{metwally2005efficient} or priority queues~\cite{pifo} requires
multiple memory accesses within the per-packet time budget. %% Recently proposed
%% advances in switch hardware may make such data structures
%% possible~\cite{pifo}.
%
However, as we show in \Sec{sec:algorithm}, we are able to adapt the key ideas
of the \spacesaving algorithm and combine it with the functionality of emerging
switch hardware to develop an effective heavy hitter algorithm.
%adapts the key ideas of
%the space saving algorithm and leverages emerging switch hardware with simple
%data structures.

%% file: algorithm.tex
\newcommand{\baselineAlgorithm}{%
\begin{algorithm}
\SetAlgoLined
$i = 0$\;
\lnl{loop} \While{i $<$ d}{
$l = h_i(ikey)$\;
\uIf{$key_l = ikey$}{$val_l$++; end\;}
\uElseIf{$val_l = 0$}{l = (ikey, 1); end\;}
\Else{$min = update(min)$\;}
$i$++\;
}
\uIf{$i = d$}{$min = (ikey, val_{min} + 1)$\;}
\end{algorithm}}

\newcommand{\OurAlgorithm}{%
\begin{algorithm}
\SetAlgoLined
$i = 0, cKey = iKey, cVal = 1$\;
\lnl{loop} \While{i $<$ d}{
$l = h_i(cKey)$\;
\uIf{$key_l = ckey$}{$val_l$ += $cVal$; end\;}
\uElseIf{$val_l = 0$}{l = (ckey, cVal); end\;}
\uElseIf{$val_l < cVal$}{swap $(key_l, val_l)$ with $(cKey, cVal)$\;}
$i$++\;
}
\end{algorithm}}

\section{HashPipe Algorithm}\label{sec:algorithm} 

As described in \Sec{sec:related}, \TheSystem is heavily inspired by the \spacesaving algorithm \cite{metwally2005efficient}, which we describe now
(\Sec{sec:spacesaving}). In the later subsections
(\Sec{sec:sampling-d}-\Sec{sec:feed-forward}), we describe our modifications to
the algorithm to make it amenable to switch implementation.

\subsection{\SpaceSaving\ Algorithm}
\label{sec:spacesaving}

To track the $k$ heaviest items, \spacesaving uses a table with $m$ (which is
$O(k)$) slots, each of which identifies a distinct flow key and its counter. All
slots are initially empty, \ie keys and counters are set to zero.

\begin{algorithm}
\DontPrintSemicolon % Some LaTeX compilers require you to use
% \dontprintsemicolon instead
\Comment{Table $T$ has $m$ slots, either containing $(key_j, val_j)$ at slot $j
  \in \{1, \ldots, m\}$, or empty. Incoming packet has key $iKey$.}\;

\eIf{$\exists$ slot $j$ in $T$ with $iKey = key_j$}{
  $val_j \gets val_j + 1$\;
}{
  \eIf{$\exists$ empty slot $j$ in $T$}{
    $(key_j, val_j) \gets (iKey, 1)$\;
  }{
    $r \gets argmin_{j \in \{1,\ldots,m\}}(val_j)$\;\label{alg:ss:minfind}
    $(key_r, val_r) \gets (iKey, val_r+1)$\;
  }
}
%% $min \gets 1$\;
%% \For{$i \gets 2$ \textbf{to} $m$} {
%%   \If{$val_i < val_{min}$} {
%%     $min \gets i$\;
%%   }
%% }
%% $min \gets (iKey, val_{min} + 1)$\;
\caption{\spacesaving algorithm~\cite{metwally2005efficient}}
\label{algo:spacesaving}
\end{algorithm}

The algorithm is summarized in \Alg{spacesaving}.\footnote{We show the algorithm
  for packet counting; it easily generalizes to byte counts.}
When a packet arrives, if its corresponding flow isn't already in the table, and
there is space left in the table, \spacesaving inserts the new flow with a count
of 1. If the flow is present in the table, the algorithm updates the
corresponding flow counter. However, if the table is full, and the flow isn't
found in the table, the algorithm replaces the flow entry that has the {\em
  minimum} counter value in the table with the incoming packet's flow, and
increments this minimum-valued counter.

This algorithm has three useful accuracy properties~\cite{metwally2005efficient}
that we list here. Suppose the true count of flow $key_j$ is $c_j$, and its
count in the table is $val_j$. First, no flow counter in the table is ever
underestimated, \ie $val_j \geq c_j$.
%%  since counters are never
%% decremented and consequently, the minimum is increasing.
Second, the minimum value in the table $val_r$ is an upper bound on the
overestimation error of any counter, \ie $val_j \leq c_j + val_r$.
%% This is because when a flow's counter is initialized with $val_m + 1$, its
%% value is overestimated by at most $val_m$ (if it was a brand new flow that
%% should have otherwise had a counter value $1$).
Finally, any flow with true count higher than the average table count, \ie $c_j
> C/m$, will always be present in the table (here $C$ is the total packet count
added into the table). This last error guarantee is particularly useful for the
threshold-heavy-hitter problem (\Sec{sec:problem}), since by using $1/t$
counters, we can extract all flows whose true count exceeds a threshold $t$ of
the total count. However, this guarantee cannot be extended directly to the
top-$k$ problem, since due to the heavy-tailed nature of
traffic~\cite{estan2002new}, the $k$th heaviest flow contributes nowhere close
to $1/k$ of the total traffic.

The operation of finding the minimum counter in the table (line
\ref{alg:ss:minfind})---possibly for each packet---is difficult within switch
hardware constraints (\Sec{sec:related}).
In the following subsections, we discuss how we incrementally modify the
\spacesaving algorithm to run on switches.

%% While this seems like the best way to approach this problem, this is not
%% feasible in practice since it needs one to examine the entire list (entries in
%% the form of a table) which is not feasible at line-rate. Alternatively, one
%% could use a sorted data structure like a heap that would maintain the
%% minimum. However, complex data structures like heaps or sorted lists do not
%% involve deterministically O(1) cost per packet since once in a while, the
%% compression process for preserving order could involve an O(k) operation. Hence,
%% we need a different version amenable on hardware. \jrex{nice paragraph but very
%%   terse if readers haven't seen these algorithms before. also, in the
%%   pseudocode, min is sometimes a single number and sometimes a tuple}

\subsection{Sampling for the Minimum Value}
\label{sec:sampling-d}

The first simplification we make is to look at the minimum of a small, {\em
  constant} number $d$ of randomly chosen counters, instead of the entire table
(in line \ref{alg:ss:minfind}, \Alg{spacesaving}). This restricts the worst-case
number of memory accesses per packet to a small fixed constant $d$.  The
modified version, \Baseline, is shown in \Alg{Baseline}. The main change from
\Alg{spacesaving} is in the set of table slots examined (while looking up or
updating any key), which is now a set of $d$ slots obtained by hashing the
incoming key using $d$ independent hash functions.

In effect, we {\em sample} the table to estimate a minimum using a few
locations. However, the minimum of just $d$ slots can be far from the minimum of
the entire table of $m$ slots. An inflated value of the minimum could impact the
useful error guarantees of \spacesaving (\Sec{sec:spacesaving}). %However, we
%show analytically that the overestimation error is not too far from the average
%table count $C/m$ with high probability. Further, 
Our evaluations in \Sec{subsec:comparisonIdeal} show that the distributions of the
minimum of the entire table and of the subsample are comparable.

%% In order to determine the minimum among the entries of a hash table, one would
%% usually need to scan all the complete set of key-value pairs. However, given
%% that that is not feasible, and that only reading a small constant number of
%% locations would be feasible, we attempt to approximate the minimum by reading
%% these locations alone and finding the minimum among them. Section
%% \ignore{include eval} shows that this process of approximating the minimum
%% results in an approximation that is reasonably close to the minimum of the
%% entire table. Independent hash functions are used to pick the indices from where
%% the values would be read to approximate the minimum.

%% the switch to generate multiple indices
%% in the same table by hashing on the incoming flow with $d$ different hash
%% functions. Further,

However, \Alg{Baseline} still requires the switch to read $d$ locations at once
to determine their minimum, and then write back the updated value. This
necessitates a read-modify-write operation, involving $d$ reads and one write
{\em anywhere in table memory}, within the per-packet time budget. However,
multiple reads to the same table ($d > 1$) are supported neither in switch
programming languages today~\cite{bosshart2014p4} nor in emerging programmable
switching chips~\cite{RMT}. Further, supporting multiple reads for every packet
at line rate requires multiported memories with strict access-latency
guarantees, which can be quite expensive in terms of area and
power~\cite{cmos-vlsi-design-book, vlsi-memory-lecture}.

\subsection{Multiple Stages of Hash Tables}
\label{sec:multiple-stages-d}
The next step is to reduce the number of reads to memory to facilitate operation
at line rate. We split the counter table $T$ into $d$ disjoint tables, and we
read exactly one slot per table. The algorithm is exactly the same as
\Alg{Baseline}; except that now hash function $h_i$ returns only slots in the
$i$th stage.

This design enables pipelining the memory accesses to the $d$ tables, since
different packets can access different tables at the same time. However, packets
may need to make two passes through this pipeline: once to determine the counter
with the minimum value among $d$ slots, and a second time to update that
counter. The second pass is possible through ``recirculation'' of packets
through the switching pipeline~\cite{p4-v1.1-spec, ovs-packet-recirculation,
  arista-7050-datasheet} with additional metadata on the packet, allowing the
switch to increment the minimum-valued counter in the second pass. However, the
second pass is potentially needed for {\em every packet}, and recirculating
every packet can halve the pipeline bandwidth available to process packets.

%% We essentially use $d$ different hash functions on the incoming flow’s key, one
%% per stage to pick exactly one location amongst the entries per stage. In this
%% manner, we are able to reduce the number of reads to one per-stage and carry
%% over the read values as metadata. The computation of the minimum can be done
%% using this state carried over at the end of the processing pipeline. However, if
%% the location where the new flow is to be rewritten is anything other than the
%% last stage, this would pose a problem. The feed-forward \ignore{cite}
%% restriction implies that a packet cannot make modifications to earlier stages in
%% the pipeline once it has passed through them. Hence, if this model was to work,
%% the packet would need to be recirculated into the pipeline and all incoming
%% packets would have to be stalled until the current packet finishes modification
%% to the appropriate pipelined stage. Clearly, this wouldn’t be feasible at line
%% rate where each stage processes a packet independently and atomically in a
%% pipelined fashion to speed up processing.

\begin{algorithm}
\DontPrintSemicolon % Some LaTeX compilers require you to use \dontprintsemicolon instead
%\KwIn{A finite set $A=\{a_1, a_2, \ldots, a_n\}$ of integers}
%\KwOut{The largest element in the set}
\Comment{Hash functions $h_i(iKey) \rightarrow$ slots, $i \in \{1,\dots,d\}$}\;

%% \For{$i \gets 1 $ \textbf{to} $d$} {
%% $l \gets h_i(ikey)$\;
%%   \If{$key_l = iKey$} {
%%     $val_l \gets val_l + 1$;\;
%%   }
%%   \ElseIf{$val_l = 0$}{
%%   	$(key_l,val_l) \gets (iKey, 1)$;\;
%%   }
%%   \Else{
%%   	$minLoc = update(minLoc)$
%%   }
%% }
%% \If{$i > d$}{$ minLoc \gets (ikey, val_{minLoc} + 1)$\;}

$H = \{h_1(iKey), \ldots, h_d(iKey)\}$\;
\eIf{$\exists$ slot $j \in H$ with $iKey = key_j$}{
  $val_j \gets val_j + 1$\;
}{
  \eIf{$\exists$ empty slot $j \in H$}{
    $(key_j, val_j) \gets (iKey, 1)$\;
  }{
    $r \gets argmin_{j \in H}(val_j)$\;\label{alg:hp:minfind}
    $(key_r, val_r) \gets (iKey, val_r+1)$\;
  }
}

\caption{\Baseline: Sample $d$ slots at once}
\label{algo:Baseline}
\end{algorithm}

\subsection{Feed-Forward Packet Processing}
\label{sec:feed-forward}
We now alleviate the need to process a packet more than once through the switch
pipeline, using two key ideas.

\begin{figure*}[h] 
  \begin{subfigure}[b]{0.45\linewidth}
    \centering
    \includegraphics[width=0.75\linewidth, height=2.5cm]{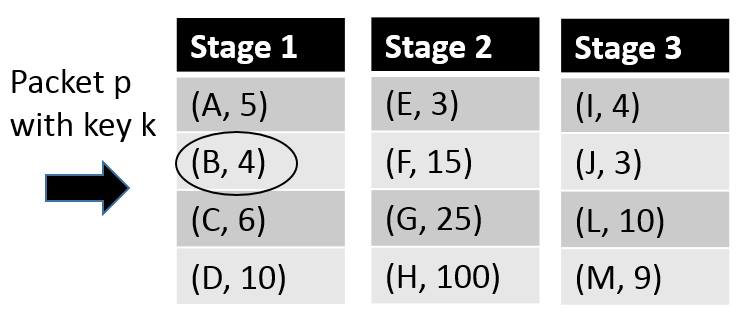} 
    \caption{Initial state of table} 
    \label{alg:a} 
    \vspace{4ex}
  \end{subfigure}%% 
  \begin{subfigure}[b]{0.45\linewidth}
    \centering
    \includegraphics[width=0.75\linewidth, height=2.5cm]{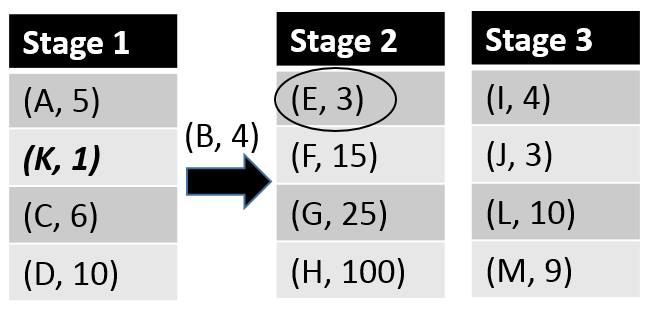} 
    \caption{New flow is placed with value $1$ in first stage}
    \label{alg:b} 
    \vspace{4ex}
  \end{subfigure} 
  \begin{subfigure}[b]{0.45\linewidth}
    \centering
    \includegraphics[width=0.75\linewidth, height=2.5cm]{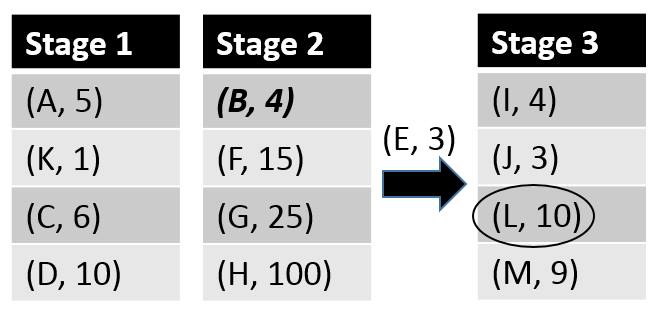} 
    \caption{$B$ being larger evicts $E$} 
    \label{alg:c} 
  \end{subfigure}%%
  \begin{subfigure}[b]{0.45\linewidth}
    \centering
    \includegraphics[width=0.75\linewidth, height=2.5cm]{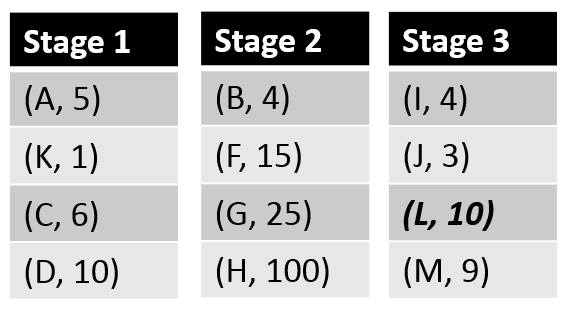} 
    \caption{$L$ being larger is retained in the table} 
    \label{alg:d} 
  \end{subfigure} 
  \\
  \vspace{0.1in}
  \caption{An illustration of \TheSystem.}
  \label{fig:HashPipe} 
\end{figure*}

\NewPara{Track a rolling minimum.} We track the minimum counter value seen so
far (and its key) as the packet traverses the pipeline, by {\em moving} the
counter and key through the pipeline as {\em packet metadata.} Emerging
programmable switches allow the use of such metadata to communicate results of
packet processing between different stages, and such metadata can be written to
at any stage, and used for packet matching at a later stage~\cite{p4-v1.1-spec}.

\begin{algorithm}
\DontPrintSemicolon % Some LaTeX compilers require you to use \dontprintsemicolon instead
%\KwIn{A finite set $A=\{a_1, a_2, \ldots, a_n\}$ of integers}
%\KwOut{The largest element in the set}
\Comment{Insert in the first stage}\;
$l_1 \gets h_1(iKey)$\;
\If{$key_{l_1} = iKey$}{
  $val_{l_1} \gets val_{l_1} + 1$\;
  end processing\;
}\ElseIf{$l_1$ is an empty slot}{
  $(key_{l_1}, val_{l_1}) \gets (iKey, 1)$\;
  end processing\;
}\Else{
  $(cKey, cVal) \gets (key_{l_1}, val_{l_1})$\;
  $(key_{l_1}, val_{l_1}) \gets (iKey, 1)$\;
}
\Comment{Track a rolling minimum}\;
\For{$i \gets 2$ \textbf{to} $d$} {
$l \gets h_i(cKey)$\;
  \If{$key_l = cKey$} {
    $val_l \gets val_l + cVal$\;
    end processing\;
  }
  \ElseIf{$l$ is an empty slot}{
    $(key_l, val_l) \gets (cKey, CVal)$\;
    end processing\;
  }
  \ElseIf{$val_l < cVal$}{
    swap $(cKey, cVal)$ with $(key_l, val_l)$
  }
}
\caption{\TheSystem: Pipeline of $d$ hash tables}
\label{algo:Sequential}
\end{algorithm}

As the packet moves through the pipeline, the switch hashes into each stage on
the {\em carried key,} instead of hashing on the key corresponding to the
incoming packet. If the keys match in the table, or the slot is empty, the
counter is updated in the usual way, and the key needs no longer to be carried
forward with the packet. Otherwise, the keys and counts corresponding to the
{\em larger} of the counters that is carried and the one in the slot is written
back into the table, and the smaller of the two is carried on the packet. We
leverage arithmetic and logical action operations available in the match-action
tables in emerging switches~\cite{RMT} to implement the counter comparison. The
key may be carried to the next stage, or evicted completely from the tables when
the packet reaches the last (\ie $d$th) stage.

\NewPara{Always insert in the first stage.} If the incoming key isn't found in
the first stage in the pipeline, there is no associated counter value to compare
with the key that is in that table. Here, we choose to {\em always insert the
  new flow} in the first stage, and evict the existing key and counter into
packet metadata. After this stage, the packet can track the rolling minimum of
the {\em subsequent stages} in the usual way described above. The final
algorithm, \TheSystem, is shown in \Alg{Sequential}.

One consequence of always inserting an incoming key in the first stage is the
possibility of duplicate keys across different tables in the pipeline, since the
key can exist at a later stage in the pipeline. Note that this is unavoidable
when packets only move once through the pipeline. It is possible that such
duplicates may occupy space in the table, leaving fewer slots for heavy flows,
and causing evictions of heavy flows whose counts may be split across the
duplicate entries.

\label{sec:coalescing}
However, many of these duplicates are easily {\em merged} through the algorithm
itself, \ie the minimum tracking merges counters when the carried key has a
``hit'' in the table. Further, switches can easily estimate the flow count
corresponding to any packet in the data plane itself by summing all the matching
flow counters; so can a data collector, after reading the tables out of the
switch. We also show in our evaluations (\Sec{subsec:sensitivity}) that
duplicates only occupy a small portion of the table memory.
%%only have a small effect on the memory available to heavy keys in the table.

\Fig{HashPipe} illustrates an example of processing a packet using \TheSystem. A
packet with a key $K$ enters the switch pipeline (a), and since it isn't found
in the first table, it is inserted there (b). Key $B$ (that was in the slot
currently occupied by $K$) is carried with the packet to the next stage, where
it hashes to the slot containing key $E$. But since the count of $B$ is larger
than that of $E$, $B$ is written to the table and $E$ is carried out on the
packet instead (c). Finally, since the count of $L$ (that $E$ hashes to) is
larger than that of $E$, $L$ stays in the table (d). The net effect is that the new
key $K$ is inserted in the table, and the minimum of the three keys $B$, $E$, and
$L$---namely $E$---is evicted in its favor.

%% file: prototype.tex
\section{HashPipe Prototype in P4}
\label{sec:prototype}

%- add details of p4 prototype
%- code fragments?
%- some details of barefoot compiler; and configurations there.
%- p4 to OVS compiler?

We built a prototype of \TheSystem using P4 version 1.1~\cite{p4-v1.1-spec}.  We
verified that our prototype produced the same results as our HashPipe simulator
by running a small number of artificially generated packets on the switch
behavioral model \cite{p4-bm} as well as our simulator, and ensuring that the
hash table is identical in both cases at the end of the measurement interval.

\begin{figure}
 \begin{lstlisting}[basicstyle=\footnotesize, caption=
  HashPipe stage with insertion of new flow. Fields prefixed with m are metadata fields., label=q1:p4-code, captionpos=b, basicstyle=\footnotesize, breaklines = true,
numbers=left, xleftmargin=2em,frame=single,framexleftmargin=2.0em]
 action doStage1(){
   mKeyCarried = ipv4.srcAddr;
   mCountCarried = 0;
  modify_field_with_hash_based_offset(mIndex, 0, stage1Hash, 32);

   // read the key and value at that location
   mKeyTable = flowTracker[mIndex];
   mCountTable = packetCount[mIndex];
   mValid = validBit[mIndex];

   // check for empty location or different key
   mKeyTable = (mValid == 0) ? mKeyCarried : mKeyTable;
   mDif = (mValid == 0) ? 0 : mKeyTable -  mKeyCarried;

   // update hash table
   flowTracker[mIndex] = ipv4.srcAddr;
   packetCount[mIndex] = (mDif == 0) ? mCountTable+1: 1;
   validBit[mIndex] = 1;

   // update metadata carried to the next table stage
   mKeyCarried = (mDif == 0) ? 0 : mKeyTable;
   mCountCarried = (mDif == 0) ? 0 : mCountTable;  
}
 \end{lstlisting}
 \end{figure}

At a high level, \TheSystem uses a match-action stage in the switch pipeline for
each hash table. In our algorithm, each match-action stage has a single default
action---the algorithm execution---that applies to every packet. Every stage
uses its own P4 \emph{register arrays}---stateful memory that persists across
successive packets---for the hash table. The register arrays maintain the flow
identifiers and associated counts.
%% In this context, every stage's P4
%% registers act as a hash table where the flow identifiers and the associated
%% counts are maintained. 
The P4 action blocks for the first two stages are
presented in Listings \ref{q1:p4-code} and \ref{q2:p4-code}; actions for
further stages are identical to that of stage $2$. The remainder
of this
section walks through the P4 language constructs with specific references to
their usage in our \TheSystem prototype.\\

\noindent \textbf{Hashing to sample locations:} The first step of the action is
to hash on the flow identifier (source IP address in the listing) with a custom
hash function, as indicated in line $4$ of Listing \ref{q1:p4-code}. The result
is used to pick the location where we check for the key. The P4
behavioral model \cite{p4-bm} allows customized hash function definitions. We
use hash functions of the type $(a_i\cdot x + b_i)\%p$ where the chosen $a_i,
b_i$ are co-prime to ensure independence of the hash functions across
stages. Hash functions of this sort are implementable on hardware and have
been used in prior work~\cite{univmon,li2016flowradar}.
%% If the key is absent,
%% the flow is used as one of the samples for approximating the minimum. 

\noindent \textbf{Registers to read and write flow statistics:} The flows are
tracked and updated using three registers: one to track the flow identifiers,
one for the packet count, and one to test validity of each table index. The
result of the hash function is used to index into the registers for reads and
writes. Register reads occur in lines $6$-$9$ and register writes occur in lines
$15$-$18$ of Listing \ref{q1:p4-code}. When a particular flow identifier is read
from the register, it is checked against the one currently being
carried. Depending on whether there is a match, either the value $1$ is written
back or the current value is incremented by $1$.

\noindent \textbf{Packet metadata for tracking current minimum:} The values read
from the registers are placed in packet metadata since we cannot test conditions
directly on the register values in P4. This enables us to compute the minimum of
the carried key and the key in the table before writing back into the register
(lines $11$-$13$ of Listing \ref{q1:p4-code} and lines $3$, $6$, and $9$ of
Listing \ref{q2:p4-code}). Packet metadata also plays a crucial role in
conveying state (the current minimum flow id and count, in this case) from one
stage to
another. The metadata is later used to compute the sample minimum. The updates
that
set these
metadata across stages are similar to lines $20$-$22$ of Listing
\ref{q1:p4-code}.

\begin{figure}
 \begin{lstlisting}[basicstyle=\footnotesize, caption=
HashPipe stage with rolling minimum. Fields prefixed with m are metadata fields., label=q2:p4-code, captionpos=b, basicstyle=\footnotesize, breaklines=true,
numbers=left, xleftmargin=2em,frame=single,framexleftmargin=2.0em,escapeinside={(*}{*)}]
action doStage2{
 (*$\cdots$*)
 mKeyToWrite =  (mCountInTable <  mCountCarried) ? mKeyCarried : mKeyTable));
 flowTracker[mIndex] = (mDif == 0) ? mKeyTable : mKeyToWrite;

 mCountToWrite =  (mCountTable < mCountCarried) ? mCountCarried : mCountTable;
 packetCount[mIndex] = (mDif == 0) ? (mCountTable + mCountCarried) :  mCountToWrite;

 mBitToWrite = (mKeyCarried == 0) ? 0 : 1);
 validBit[mIndex] = (mValid == 0) ? mBitToWrite : 1);
 (*$\cdots$*)
}
\end{lstlisting}
\end{figure}

\noindent \textbf{Conditional state updates to store local maximum:} The first
pipeline stage involves a conditional update to a register to distinguish a hit
and a miss for the incoming packet key in the table (line 17,
Listing~\ref{q1:p4-code}).
Subsequent stages must also write back different values to the table key and
count registers, depending on the result of the lookup (hit/miss) and a
comparison of the flow counts.
Accordingly, we perform a conditional write into the flow id and count registers
(lines 4 and 7, Listing~\ref{q2:p4-code}).
%
%This ensure than an update happens if and only if the flow in our metadata is
%larger than the flow currently in the table, the flow matches the entry in the
%table or if the location in the table is currently empty.
%
Such conditional state updates are feasible at line rate~\cite{domino,minions}.

%% file: evaluation.tex
\section{Evaluation}
\label{sec:evaluation}
%% \jrex{start with a brief roadmap of the section.}

We now evaluate \TheSystem through trace-driven simulations. We tune the
main parameter of \TheSystem---the number of table stages $d$---in
\Sec{subsec:sensitivity}. We evaluate the performance of \TheSystem in isolation
in \Sec{subsec:isolatedEvaluation}, and then compare it to prior sampling and
sketching solutions in \Sec{subsec:comparisonRelated}. Then, we examine the
performance of \TheSystem in context of the idealized algorithms it is derived from
(\Sec{sec:algorithm}) in \Sec{subsec:comparisonIdeal}.

\NewPara{Experiment setup.} We compute the $k$ heaviest flows using two sets of
traces. The first trace is from a 10Gb/s ISP backbone link, recorded in 2016 and
available from CAIDA~\cite{caida}. We measure heavy hitters aggregated by
transport 5-tuple.
The traffic trace is 17 minutes long, and
contains 400 million packets. We split this trace into 50 chunks, each being
about 20 seconds long, with 10 million packets. The chunks on average
contain about 400,000 5-tuple flows each. Each chunk is one trial, and each data
point in the graphs for the ISP trace reports the average across 50 trials. We
assume that the switch zeroes out its tables at the end of each trial, which
corresponds to a 20 second ``table flush'' period. %% It is feasible to flush the
%% table this often in switch hardware; switches can already rewrite tables at
%% intervals of tens of seconds or smaller, \eg~\cite{bgp-neighbor-timer,
%%   netflow-timer, li2016flowradar}.

The second trace, recorded in 2010, is from a data center~\cite{theo-dc-traffic}
and consists of about 100 million packets in total. We measure heavy hitters
aggregated by source and destination IPs. We split the trace into 1 second
intervals corresponding to the time scale at which data center traffic exhibits
stability~\cite{microTE}.

The data center trace is two and a half hours long, with roughly 10K packets
(300 flows) per 
second. We additionally replay the trace at two higher packet rates to test
whether \TheSystem can provide good accuracy over the 1 second time scale. We
assume a ``typical'' average packet size of 850 bytes and a 30\% network
utilization as reported in prior traffic studies~\cite{theo-dc-traffic}. For a
switch clocked at 1GHz with 48 ports of 10Gb/s each, this corresponds to roughly
20 million packets per second through the entire switch, and 410K packets per
second through a single link. At these packet rates, the trace contains
20,000 and 3200 flows per second respectively. For each packet rate, we report
results averaged
from multiple trials of 1 second each.

%% The results we report are averaged from multiple
%% samples corresponding to 1 second chunks of the trace corresponding to the three
%% different packet rates.

%% Given that the traffic is relatively stable over 1 -
%% 2s intervals, we chunk the trace accordingly. This results in about 10K packets
%% per second. We also "fast-forward" the traces to the single switch rate of about
%% 20 million packets per second and the single link rate of 410 million packets
%% per second. Once the origincal trace is chunked accordingly, we treat each chunk
%% as one trial and each data point in the following graphs reports the average
%% across all trials.

\NewPara{Metrics.} As discussed in \Sec{sec:problem}, we evaluate schemes on
false negatives (\% heavy flows that are not reported), false positives (\%
non-heavy flows that are reported), and the count estimation error (\% error for
heavy flows). Note that for the top-$k$ problem, the false positive error is
just a scaled version of the false negative.

\subsection{Tuning \TheSystem}
\label{subsec:sensitivity}

Given a total memory size $m$, \TheSystem's only tunable parameter is the number
of table stages $d$ that it uses. Once $d$ is fixed, we simply partition the
available memory equally into $d$ hash tables. As $d$ increases, the number of
table slots over which a minimum is computed increases, leading to increased
retention of heavier keys. However, with a fixed total memory, an increase in
the value of $d$ decreases the per-stage hash table size, increasing the
likelihood of hash collisions, and also of duplicates
(\Sec{sec:feed-forward}). The switch hardware constrains the number of table
stages to a small number, \eg 6-16~\cite{intel-fm6000, RMT}.

\begin{figure}
  \centering
  \[
  \begin{array}{ccc}
	\includegraphics[width=.49\columnwidth]{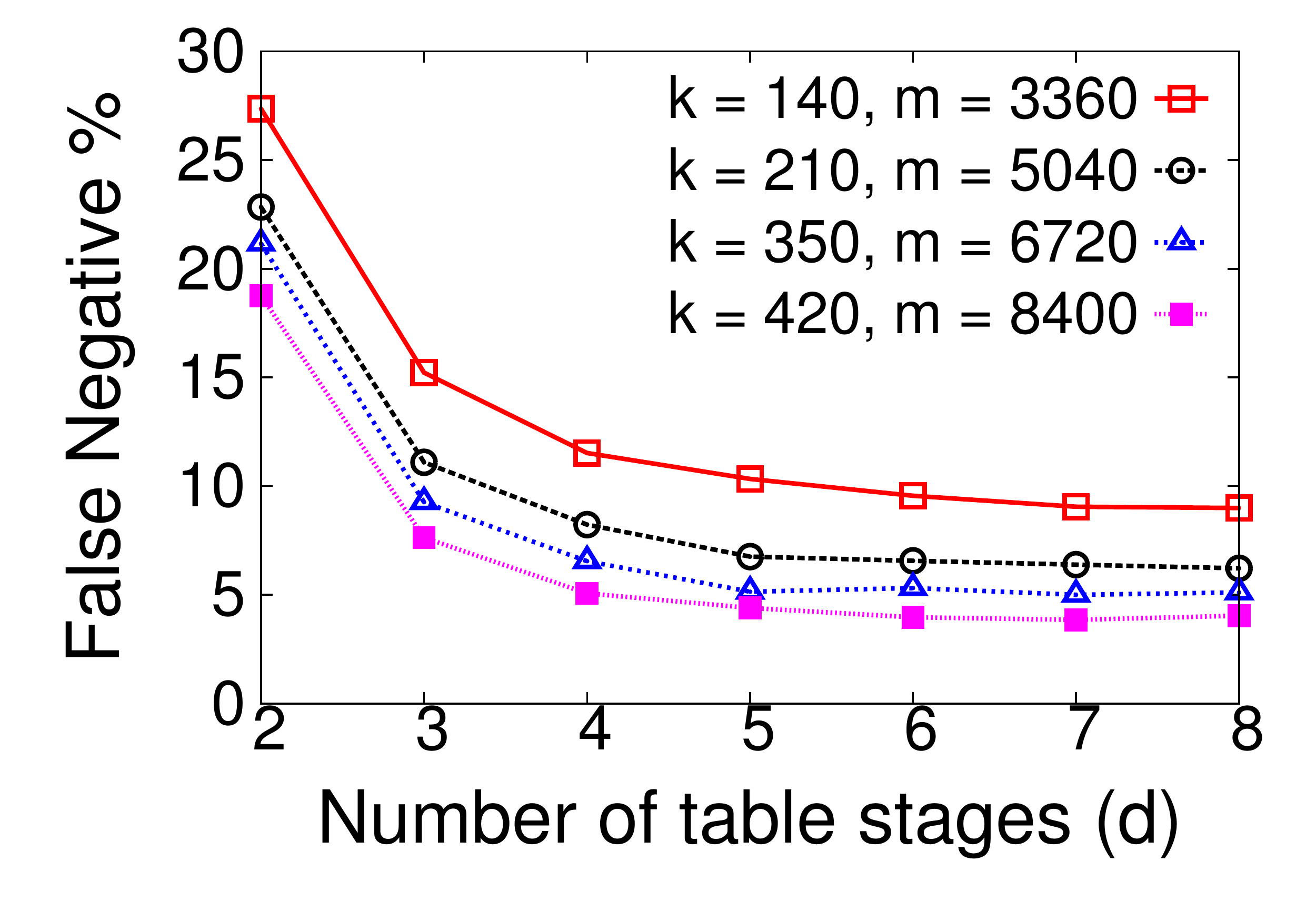} &
	\includegraphics[width=.49\columnwidth]{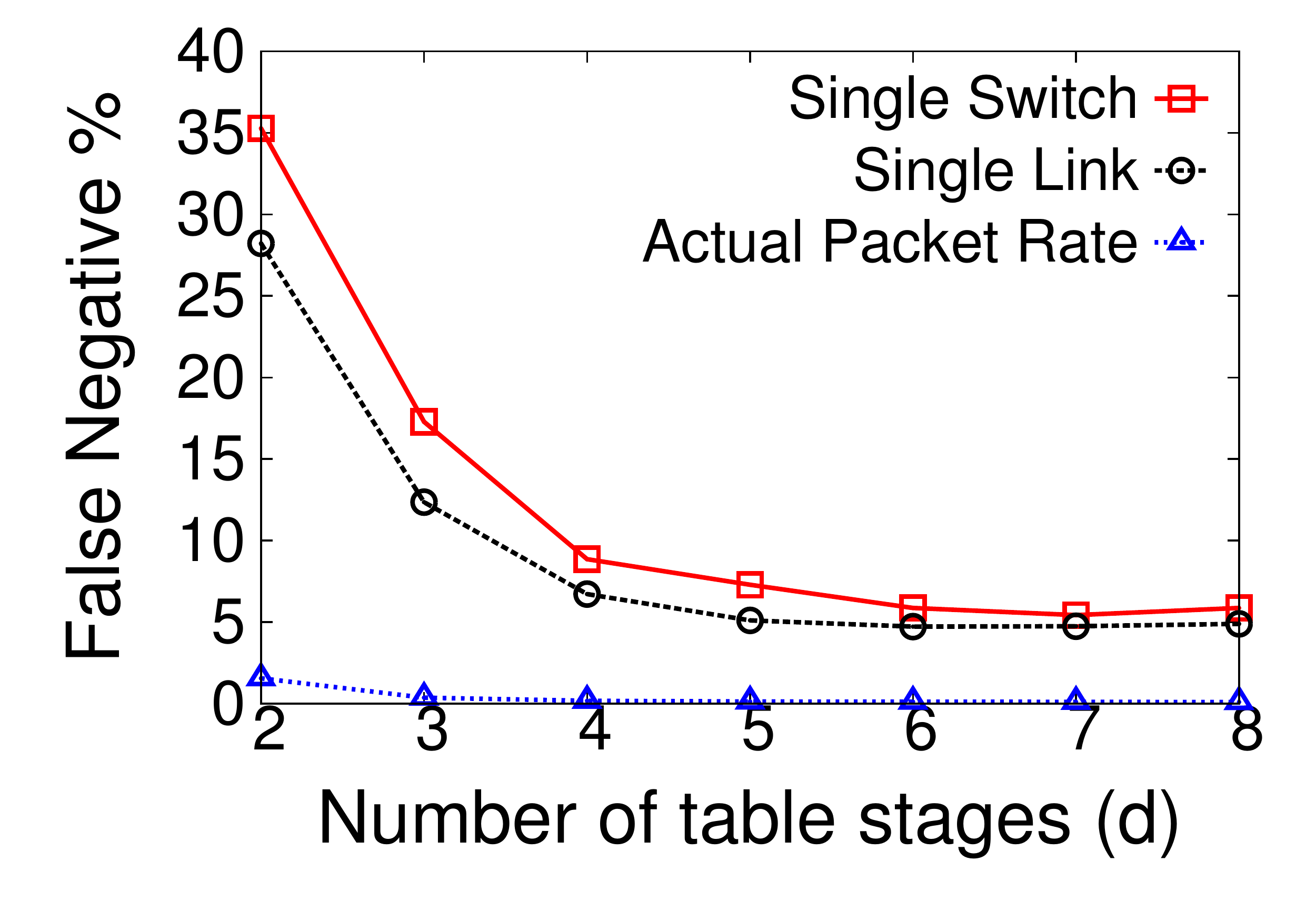}
    \\
    \mbox{(a) ISP backbone} & \mbox{(b) Data center} \\
  \end{array}
  \]
\caption{Impact of table stages ($d$) on false negatives. Error
  decreases as $d$ grows, and then flattens out.}
\label{fig:falseNegvsD}
\end{figure}

\Fig{falseNegvsD} shows the impact of changing $d$ on the false negatives. For
the ISP trace, we plot false negatives for different sizes of memory $m$ and
different number of desired heavy hitters $k$. As expected, the false negatives
reduce as $d$ increases starting at $d=2$, but the decrease quickly tapers off
in the range between 5--8, across the different $(m,k)$ curves. For the data
center trace, we show false negatives for $k=350$ with a memory of 840 counters
(15KB), and we see a similar trend across all three packet rates. The false
positives, elided here, also follow the same trend.

\begin{figure}
  \centering
  \[
  \begin{array}{ccc}
	\includegraphics[width=.49\columnwidth]{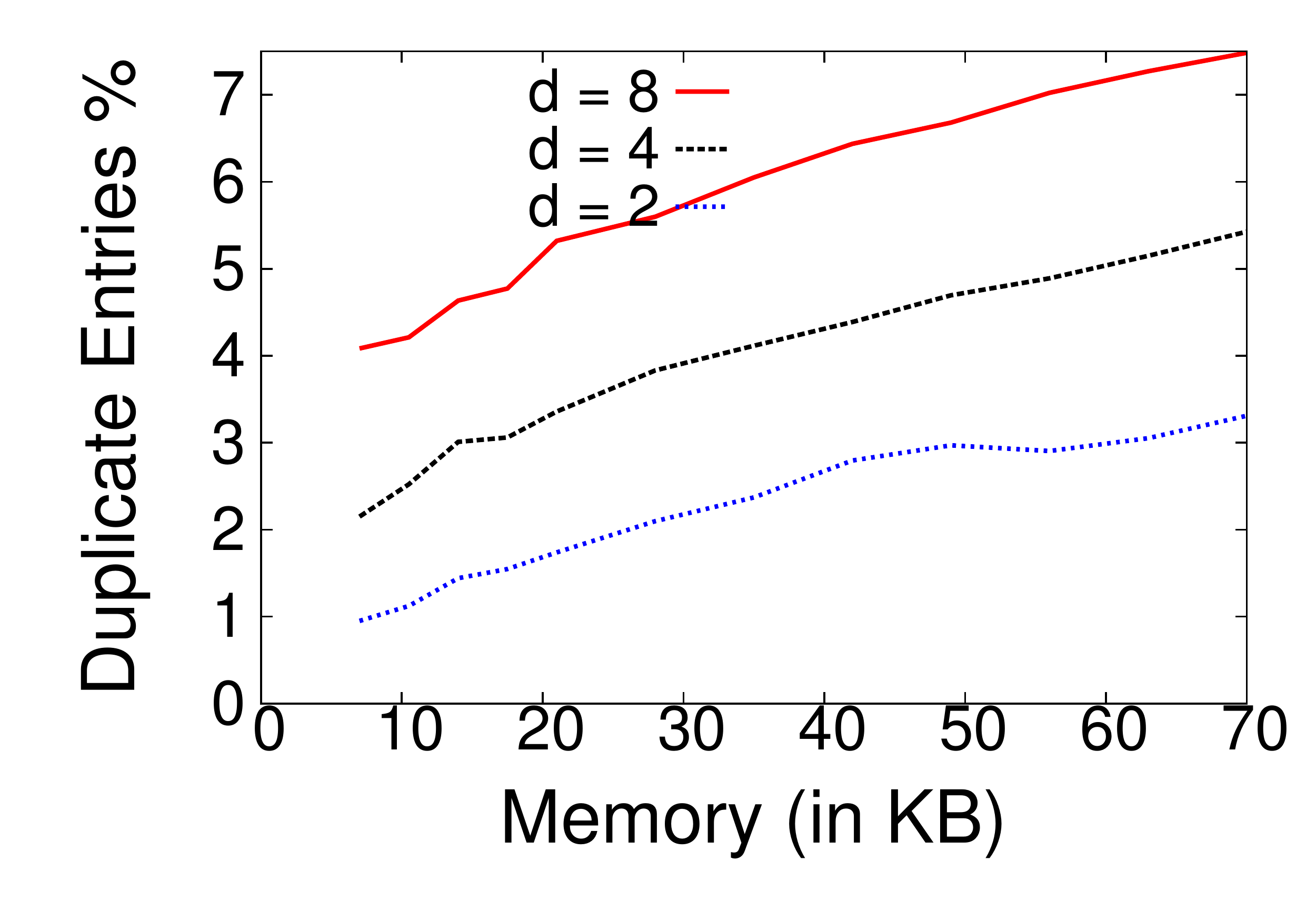} &
	\includegraphics[width=.49\columnwidth]{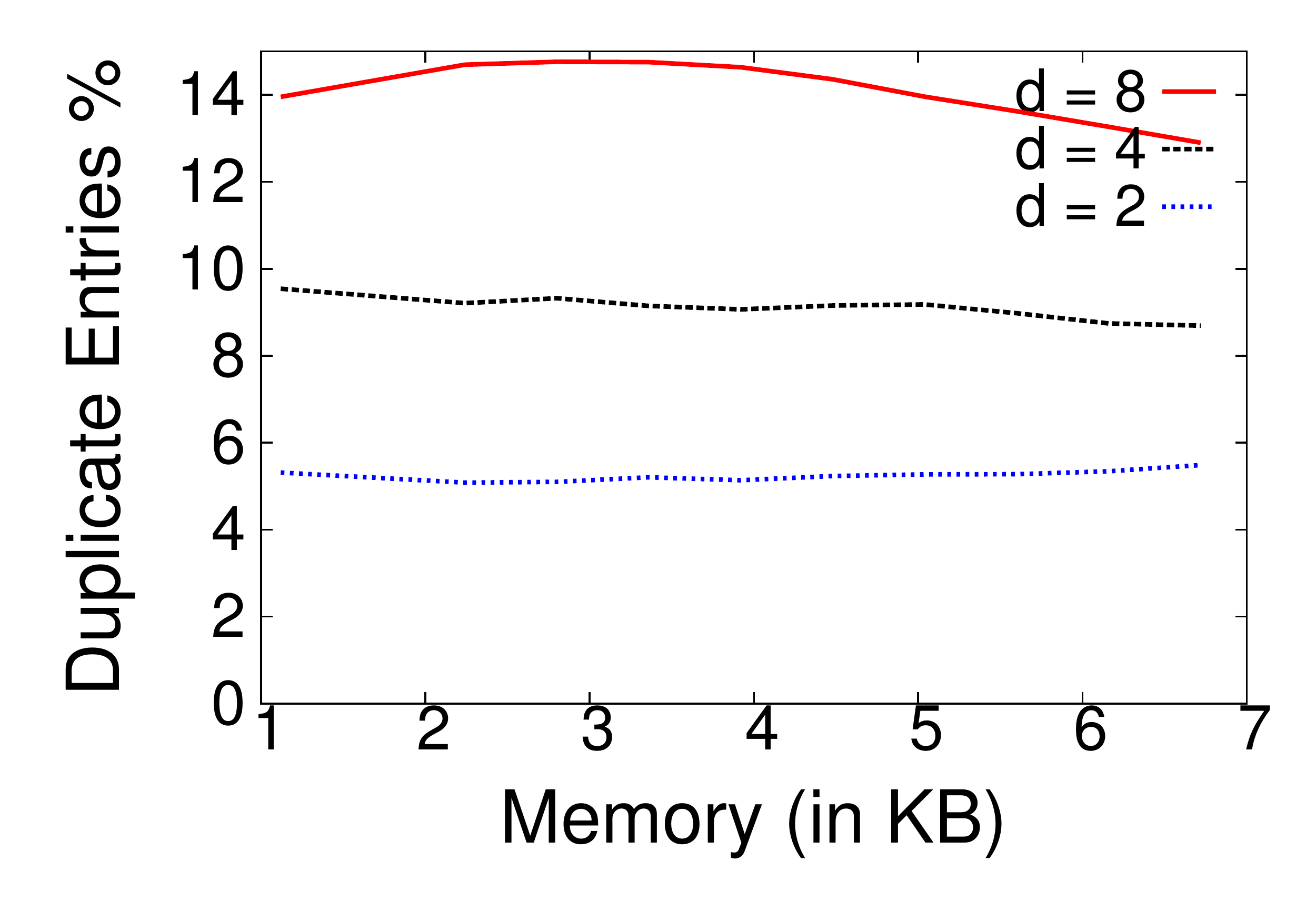}
    \\
    \mbox{(a) ISP backbone} & \mbox{(b) Data center (link)} \\
  \end{array}
  \]
\caption{Prevalence of duplicate keys in tables. For different $d$ values under
  the memory range tested, the proportion of duplicates is between 5-10\% for
  the ISP trace and 5-15\% for the data center trace (link packet rate).}
\label{fig:duplicates}
\end{figure}

To understand whether duplicates impact the false negative rates, we also show
the prevalence of duplicates in \TheSystem's hash tables in
\Fig{duplicates}. Overall, duplicates only take up between 5-10\% of the
available table size in the ISP case and between 5-14\% in the data center
case. As expected, in going from $d=2$ and $d=8$, the prevalence of duplicates
in \TheSystem's table increases.

\begin{figure}
  \centering
  \[
  \begin{array}{ccc}
	\includegraphics[width=.49\columnwidth]{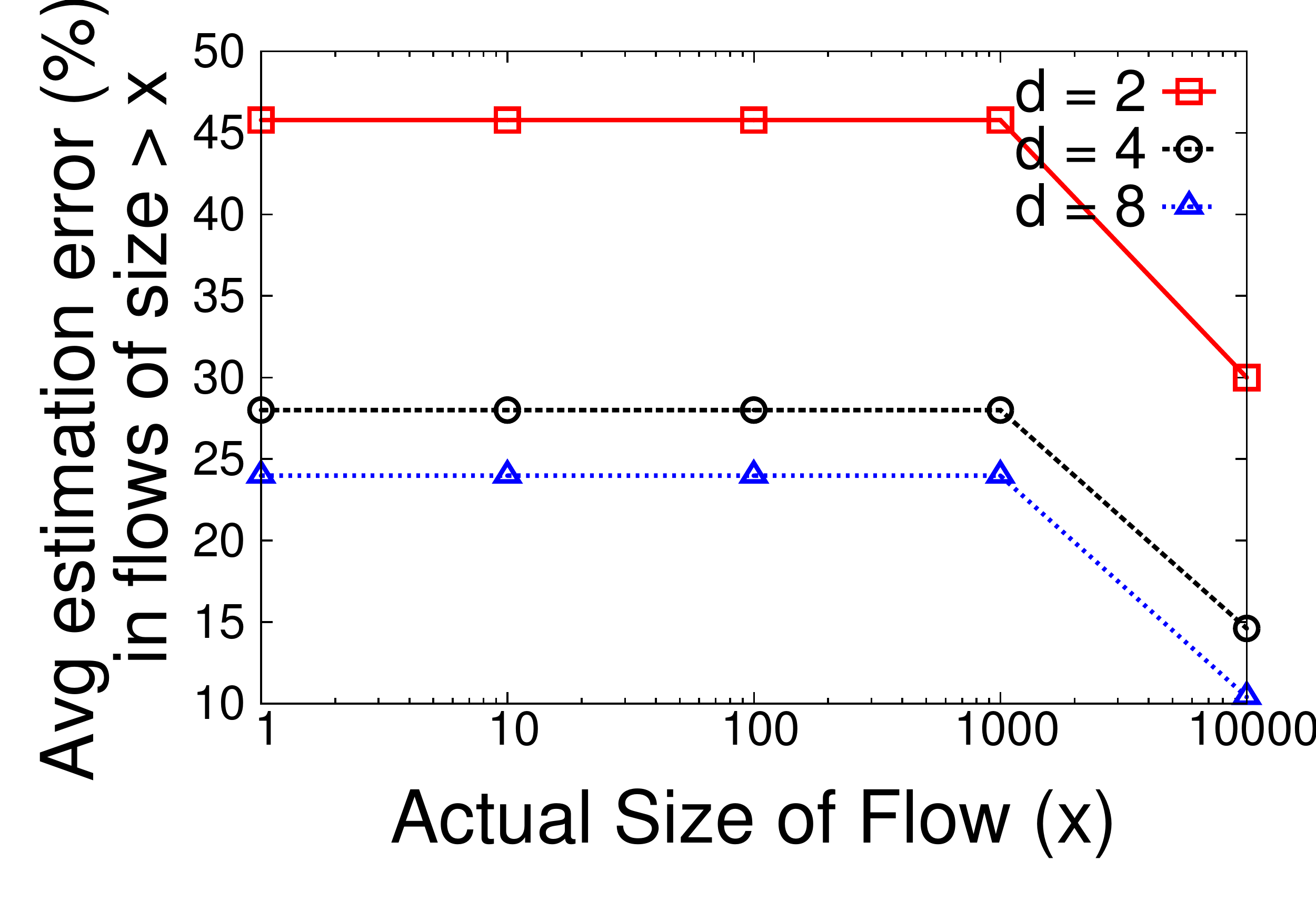} &
	\includegraphics[width=.49\columnwidth]{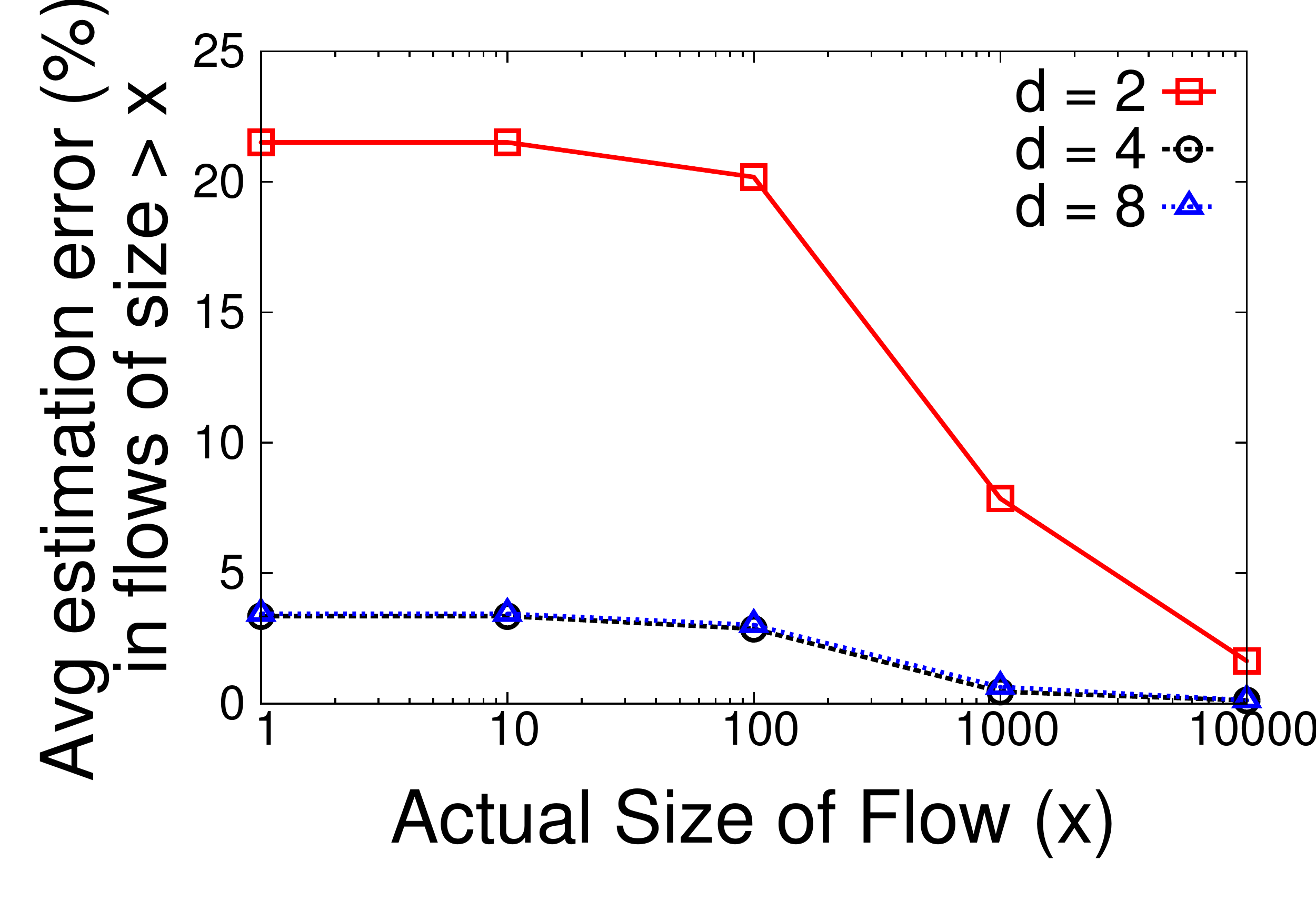}
    \\
    \mbox{(a) ISP backbone} & \mbox{(b) Data center (link)} \\
  \end{array}
  \]
\caption{Average estimation error (\%) of flows whose true counts are higher
  than the $x$-value in the graph. Error decreases as $d$ grows but the benefit
  diminishes with increasing $d$. Error decreases with actual flow size.}
\label{fig:estimation-error-D}
\end{figure}

\Fig{estimation-error-D} shows the count estimation error (as a
percentage of actual flow size) {\em for flows in
  \TheSystem's tables} at the end of each measurement interval, with a memory
size of 640 counters (11.2KB).
In general, the
error reduces as $d$ increases, but the reduction from $d=4$ to $d=8$ is less
significant than the reduction from $d=2$ to $d=4$.
In the ISP trace, the estimation error is stable across flow sizes since most
flows recorded by \TheSystem have sizes of at least 1000.
In the data center trace where there are fewer total flows, there is a more
apparent decrease in error with true flow size, with flows of size $x > 1000$
having near-perfect count estimations.

\NewPara{Choosing $d=6$ table stages.} To summarize, we find that (i) as the
number of table stages increases above $d=4$, all the accuracy metrics improve;
(ii) however, the improvement dimishes at $d=8$ and beyond, due to the increasing
prevalence of duplicates and hash collisions. These trends hold across both the
ISP and data center scenarios, for a variety of measured heavy hitters $k$ and
memory sizes. Hence, we choose $d=6$ table stages for all further experiments
with \TheSystem.

%% \iffalse
%% \begin{figure}[h]
%% \includegraphics[max height=11cm,max width=8cm]{WFalseNegsvsDsSingle.pdf}
%% \caption{Weighted fraction of Heavy Hitters Reported for Different $d$ values  with 500000 flows in total}
%% \label{fig:weightedNegvsD}
%% \end{figure}
%% \fi

%% \iffalse
%% \begin{figure}[h]
%% \includegraphics[max height=11cm,max width=8cm]{FalseNegvsDSingle.pdf}
%% \caption{Impact of the number of table stages on false negatives. Error
%%   decreases as $d$ grows, and then flattens out.}
%% \label{fig:falseNegvsD}
%% \end{figure}
%% \fi

%% \iffalse
%% \begin{figure}[h]
%% \includegraphics[max height=11cm,max width=8cm]{Duplicates.pdf}
%% \caption{Prevalence of duplicate keys in \TheSystem's tables.}
%% \label{fig:duplicates}
%% \end{figure}
%% \fi

%% \begin{figure}
%%   \centering
%%   \[
%%   \begin{array}{ccc}
%% 	\includegraphics[width=.49\columnwidth]{DsRelativeError.pdf} &
%% 	\includegraphics[width=.49\columnwidth]{TheoDsRelativeError.pdf}
%%     \\
%%     \mbox{(a)} & \mbox{(b)} \\
%%   \end{array}
%%   \]
%% \caption{Comparison of average estimation error (\%) of flows whose true counts   are higher than the $x$-value in the graph. Error
%%  decreases as $d$ grows, yet, the difference is less pronounced at higher $d$. (a) CAIDA Trace (b) Data Center packet rate for a single link}
%% \label{fig:estimation-error-D}
%% \end{figure}

%\begin{figure}[h]
%\includegraphics[max height=11cm,max width=8cm]{DsRelativeError.pdf}
%\caption{Comparison of average estimation error (\%) of flows whose true counts
%  are higher than the $x$-value in the graph. Error
% decreases as $d$ grows, yet, the difference is less pronounced at higher $d$}
%\label{fig:estimation-error-D}
%\end{figure}

%% In order to enure that the algorithm preserves the feed-forward property,
%% \TheSystem always inserts the incoming flow in stage $1$ an then computes the
%% minimum in a rolling fashion (\Sec{sec:feed-forward}). However, this could
%% induce some duplicates in the table when a flow present in a later table
%% stage is inserted afresh in the first table stage. Figure
%% \ref{fig:duplicates} shows that the fraction of the table occupied by
%% duplicates increases as the memory provisioned is increased but that this is
%% a small fraction (around $5\%$) of the total memory available for
%% heavy-hitter detection. %smaller memories, fewer %duplicates - get evicted
%% (or coalesced) for %larger flows - shows up as underestimates?

\subsection{Accuracy of \TheSystem}
\label{subsec:isolatedEvaluation}

We plot error \vs memory tradeoff curves for \TheSystem, run with $d=6$ table
stages. Henceforth, unless mentioned otherwise, we run the data center trace at
the single link packet rate.

\begin{figure}
  \centering
  \[
  \begin{array}{ccc}
	\includegraphics[width=.49\columnwidth]{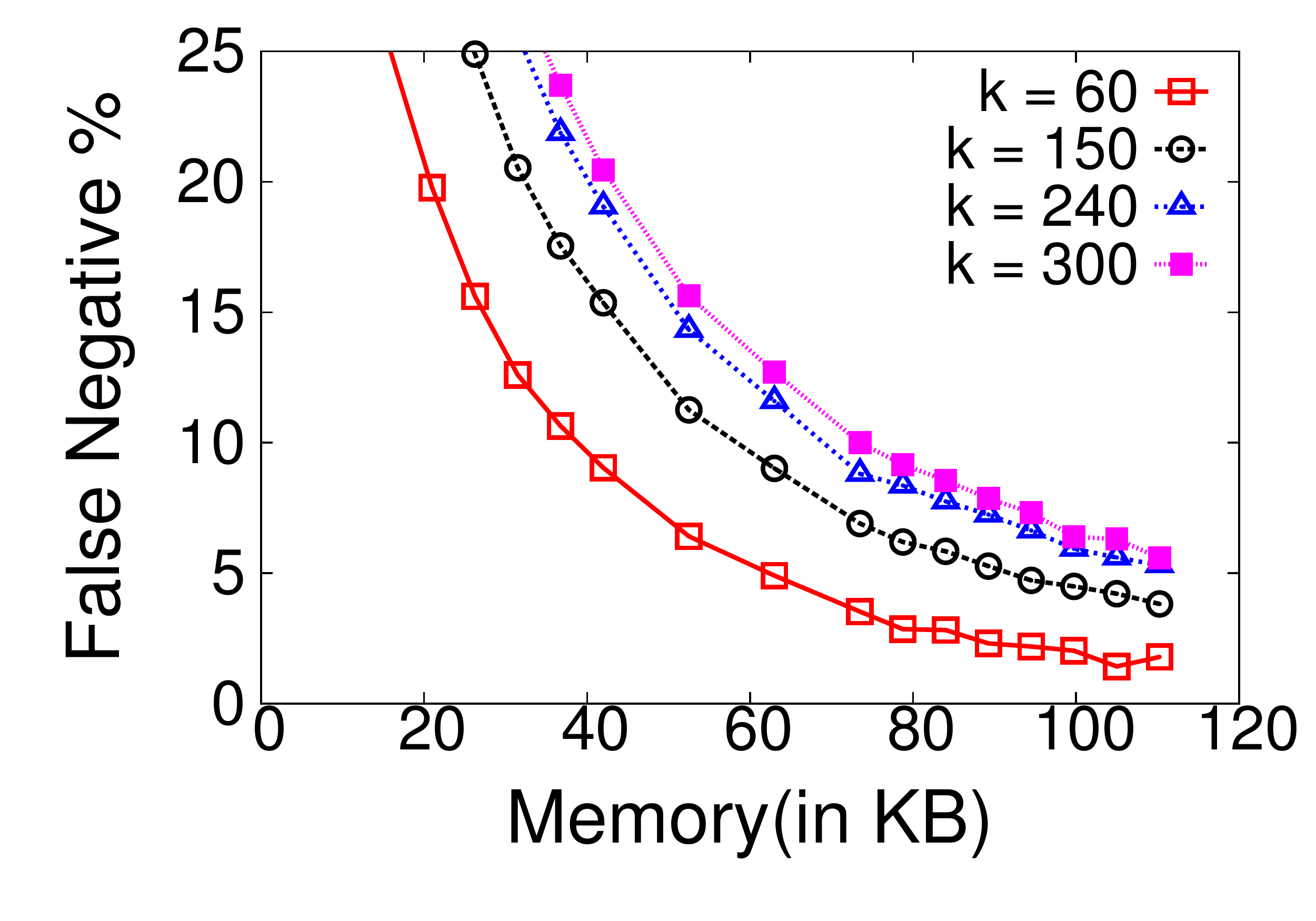} &
	\includegraphics[width=.49\columnwidth]{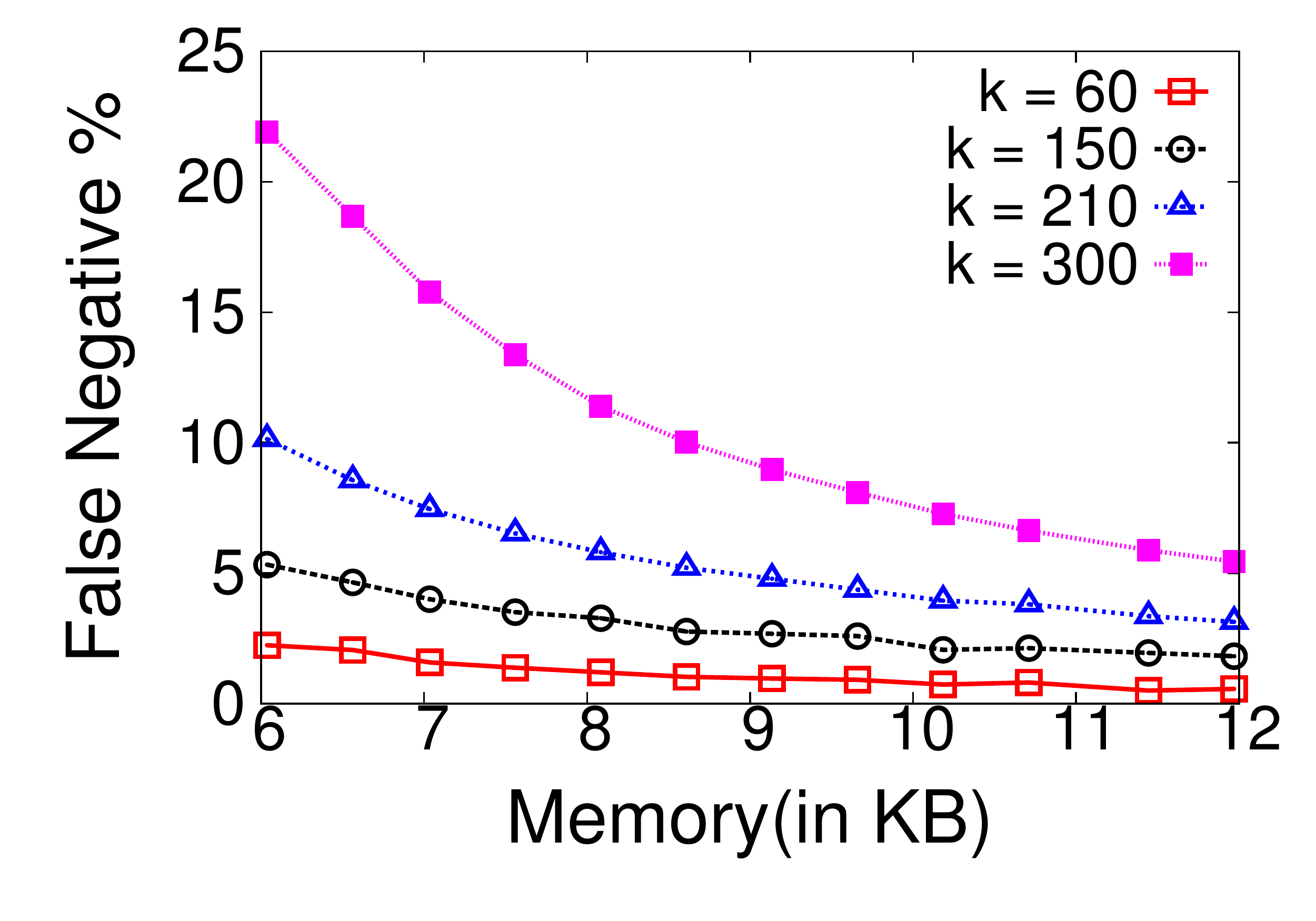}
    \\
    \mbox{(a) ISP backbone} & \mbox{(b) Data center} \\
  \end{array}
  \]
\caption{False negatives of \TheSystem with increasing memory. Each trial trace
  from the ISP backbone contains an average of 400,000 flows, yet \TheSystem
  achieves 5-10\% false negatives for the top 60-300 heavy hitters with
  just 4500 flow counters.}
\label{fig:HPNegvsM}
\end{figure}

\NewPara{False negatives.} \Fig{HPNegvsM} shows the false negatives as memory
increases, with curves corresponding to different numbers of reported heavy
hitters $k$. We find that error decreases with allocated memory across a range
of $k$ values, and settles to under 10\% in all cases on the ISP trace at 80KB
of memory, which corresponds to 4500 counters. 
In any one trial with the ISP trace, there are on average 400,000 flows, which
is two orders of magnitude higher than the number of counters we use.
In the data center trace, the error settles to under 10\% at just 9KB of memory
(520 counters) for all the $k$ values we tested.\footnote{The minimum memory
  required at $k=300$ is 6KB ($\approx$ 300 counters).}

These results also enable us to understand the interplay between $k$ and the
memory size required for a specific accuracy. For a 5\% false negative rate in the
ISP trace, the memory required for $k=60$ is 60KB ($3375 \approx 55k$ counters),
whereas the memory required for $k=300$ is 110KB ($6200 \approx 20k$ counters).
%% At 110KB memory ($m=6200$
%% counters) for the ISP trace, we find that the error is about 5\% for $k=300$
%% ($m \approx 20k$), and around 2\% for $k=60$ ($m \approx 100k$). 
%
%% For $k=300$, the error drops under 10\% at about 4000 counters ($\sim 13k$)
%% counters, and for $k=60$ around 2000 counters ($\sim 33k$).
%% (As memory increases beyond the range shown, false negatives continue to
%%drop.)
%% Recall that there are about 400,000 flows on average in any one trial in this trace, which is two orders of magnitude higher than the number of counters used here.
In general, the factor of $k$ required in the number of counters to
achieve a particular accuracy reduces as $k$ increases.

%% unsure about this figure, not turning out the way i hoped

\begin{figure}
  \centering
  \[
  \begin{array}{ccc}
	\includegraphics[width=.49\columnwidth]{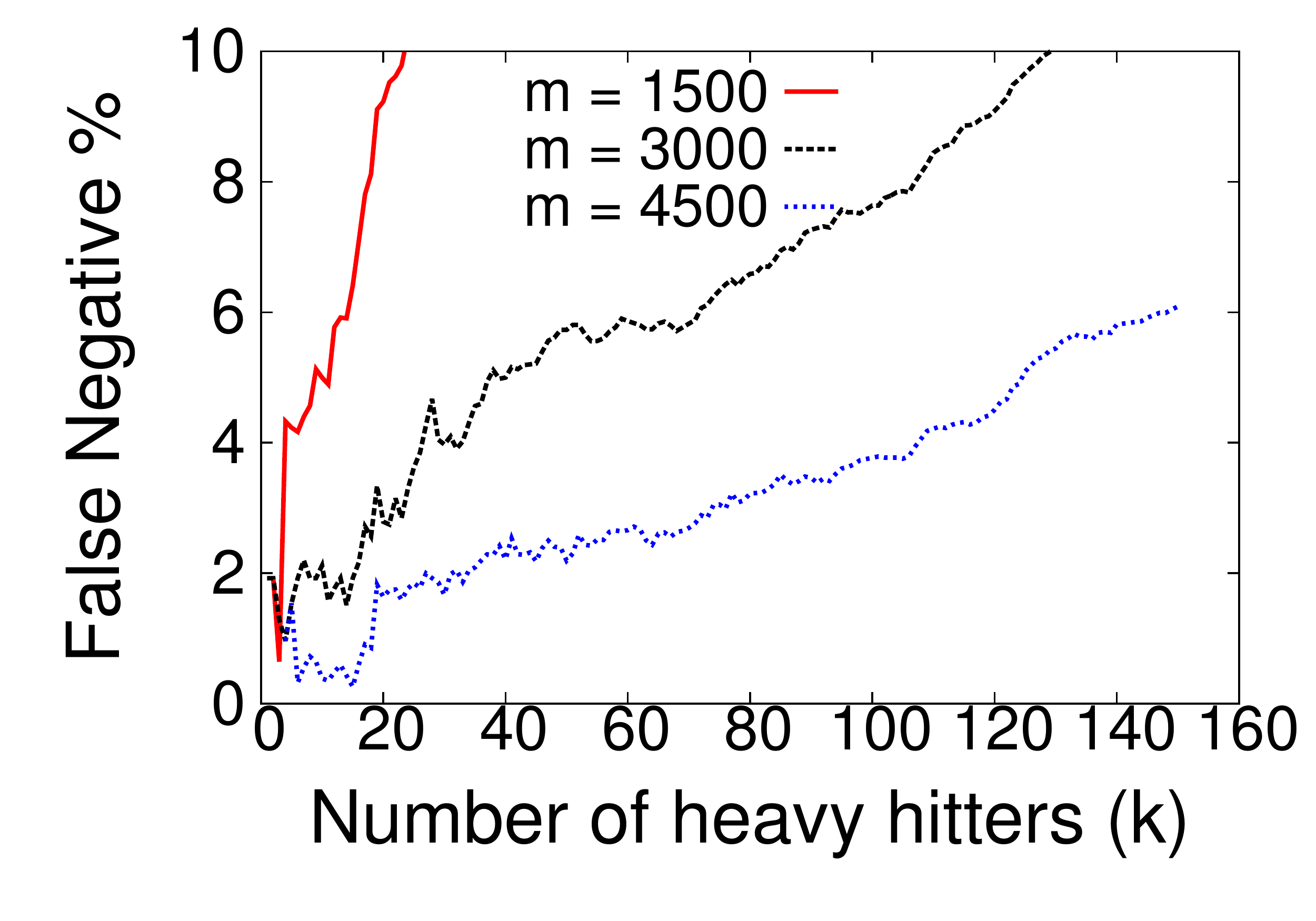} &
	\includegraphics[width=.49\columnwidth]{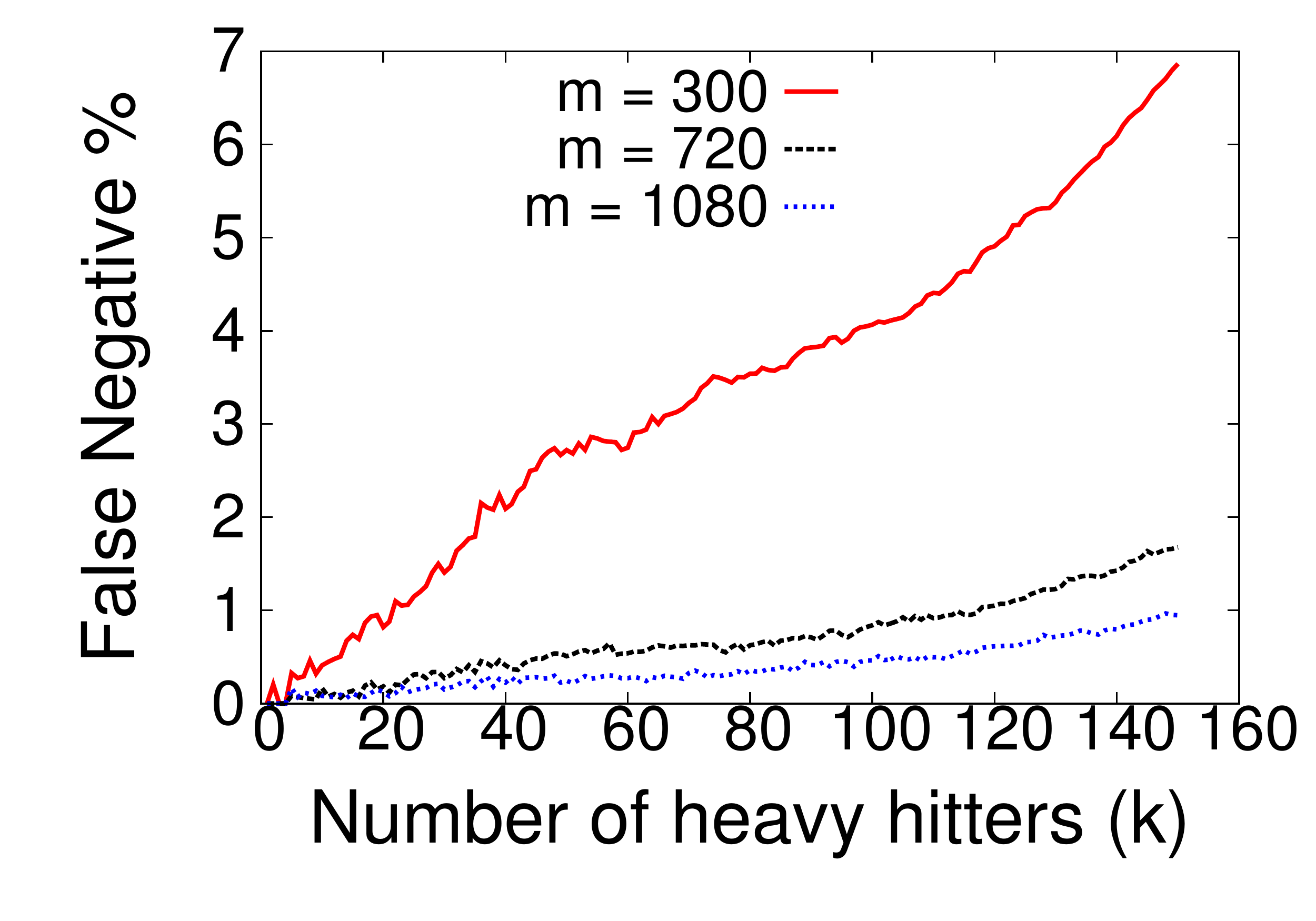}
    \\
    \mbox{(a) ISP backbone} & \mbox{(b) Data center} \\
  \end{array}
  \]
\caption{False negatives against different numbers of reported heavy flows
  $k$. The heavier a flow is, the less likely that it is missed.}
\label{fig:falseNegvsK}
\end{figure}

%rephrase that it helps understand error
\NewPara{Which flows are likely to be missed?} %% We achieve a 10\% false negative
%% rate for the top 300 flows with 80KB memory, so 
It is natural to ask which flows
are more likely missed by \TheSystem. \Fig{falseNegvsK} shows how the false
negative rate changes as the number of desired heavy hitters $k$ is varied, for
three
different total memory sizes. We find that {\em the heaviest flows are least
  likely to be missed}, \eg with false negatives in the 1-2\% range for the top
20 flows, when using 3000 counters in the ISP trace. This trend is intuitive,
since \TheSystem
prioritizes the retention of larger flows in the table
(\Sec{sec:spacesaving}). Most of the higher values of false negative errors are
at larger values of $k$, meaning that the smaller of the top $k$
flows are more likely to be missed among the reported flows.

%% While the overall false
%% negative rate that we achieve for the top-$k$ flows is around $10\%$, Figure
%% \ref{fig:falseNegvsK} indicates that if we were interested in a smaller number
%% of heavy hitters, the false negative rate would be even lower.  Given that
%% \TheSystem prioritizes heavier flows over smaller ones, as $k$ increases, the
%% false negative rate in reporting the corresponding top-$k$ heavy hitters also
%% increases.  This does indicate, though, that the majority of the errors in
%% \ref{fig:HPPosvsM} (\ref{fig:HPNegvsM}) and \ref{fig:falseNegvsD} are arising
%% from the smaller flows and the heavier flows are retained with very high
%% probability.
% typical number that might be useful for applications?
%% anything more?

% \begin{figure}[ht]
% \includegraphics[width=.49\columnwidth]{FalseNegvsKSingle.pdf}
% \includegraphics[width=.49\columnwidth]{FalseNegvsKTheo.pdf}
% \caption{False negatives against different numbers of reported heavy flows
%   $k$. The heavier a flow is, the less likely it is missed.}
% \label{fig:falseNegvsK}
% \end{figure}

\begin{figure}
  \centering
  \[
  \begin{array}{ccc}
	\includegraphics[width=.49\columnwidth]{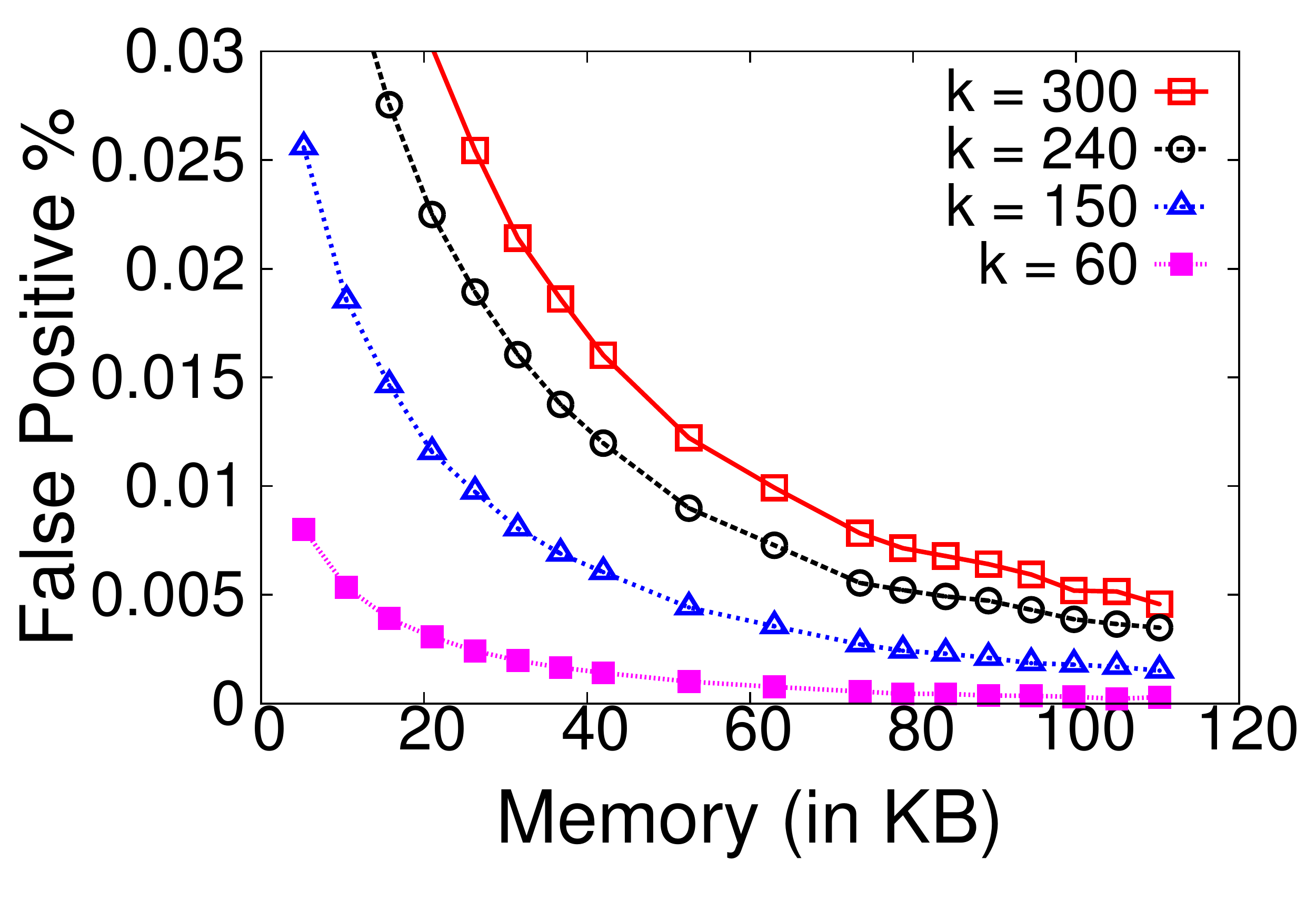} &
	\includegraphics[width=.49\columnwidth]{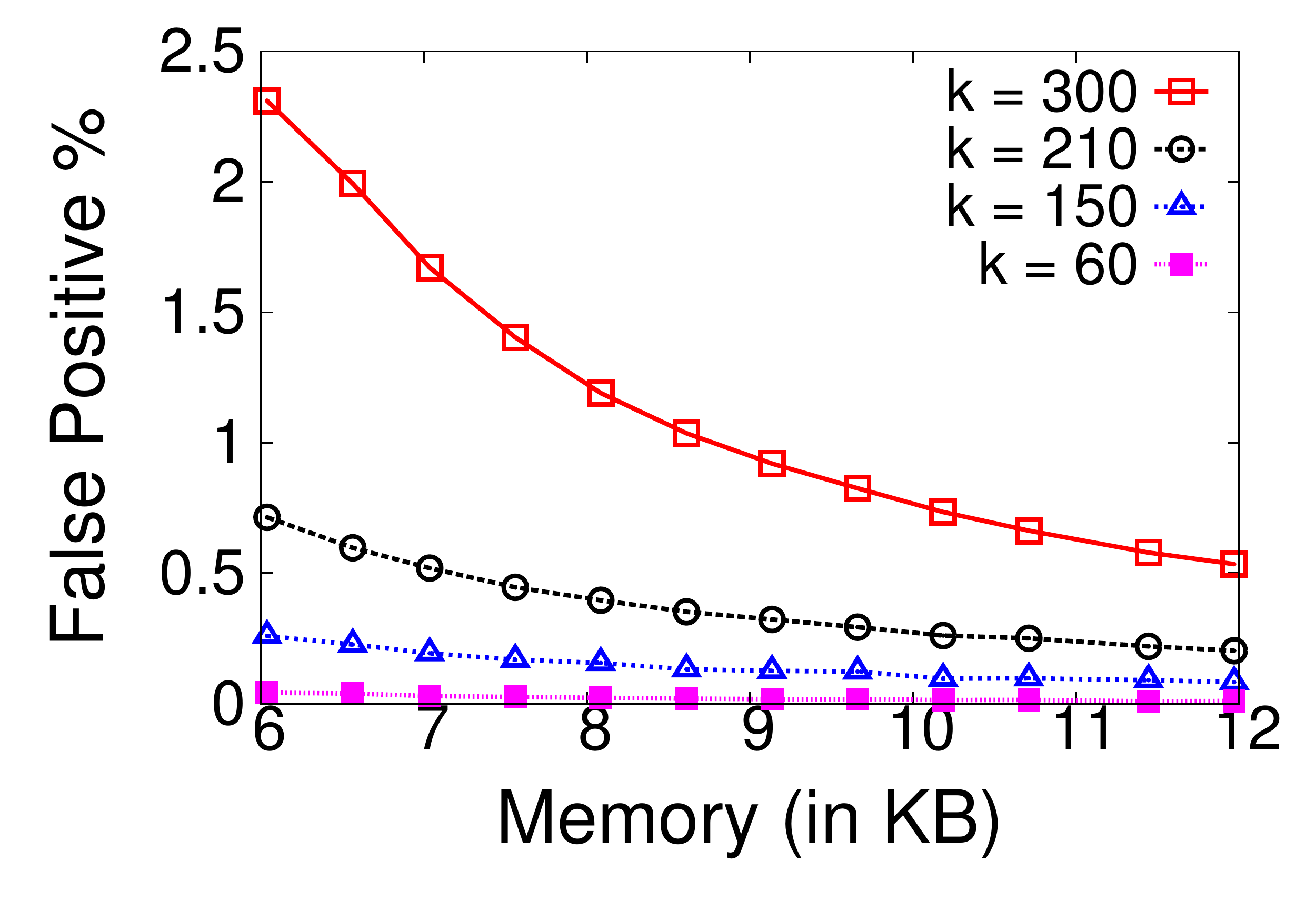}
    \\
     \mbox{(a) ISP backbone} & \mbox{(b) Data center} \\
  \end{array}
  \]
\caption{\TheSystem has false positive rates of 0.01\%-0.1\% across a range of
  table sizes and reported heavy hitters in the ISP trace, and under 3\% in
  the data center trace.}
\label{fig:HPPosvsM}
\end{figure}

\NewPara{False positives.} \Fig{HPPosvsM} shows the false positives of
\TheSystem against varying memory sizes, exhibiting the natural trend of
decreasing error with increasing memory. In particular, we find that with the
ISP trace, false positive rates are very low, partly owing to the large number
of small flows. On all curves, the false positive rate is smaller than 0.1\%,
dropping to lower than 0.01\% at a table size of 80KB. In the data center trace,
we find that the false positive rate hovers under 3\% over a range of
memory sizes and heavy hitters $k$.
%\ngs{Perhaps add counter estimation errors.}

%% \iffalse
%% \begin{figure}[ht]
%% \includegraphics[width=\columnwidth]{FalsePosvsMsSingle.pdf}
%% \caption{\TheSystem has false positive rates of 0.01\%-0.1\% across a range of
%%   table sizes and reported heavy hitters in the ISP trace, and under 1-2\% in
%%   the data center trace.}
%% \label{fig:HPPosvsM}
%% \end{figure}
%% \fi

%Choose parameters for good performance of our scheme. Given a K (for top K) or threshold (for threshold HH), how do you determine
%- number of table stages? 
%- assymetry - wanna explore this at all?
%- memory size?

%(2) Memory/other overheads - get away with not keeping entire key if the goal
%isn't to report to the controller, but instead to act on hh directly in the
%dataplane

In summary, \TheSystem performs well both with the ISP backbone and data center
traces, recognizing heavy-hitter flows in a timely manner, \ie within 20 seconds
and 1 second respectively, directly in the data plane.

\subsection{\TheSystem\ \vs Existing Solutions}\label{subsec:comparisonRelated}

\NewPara{Comparison baselines.} We compare \TheSystem against two
schemes---representative of sampling and sketching---which are able to estimate
counts in the switch directly (\Sec{sec:related}). We use the same total memory
as earlier, and measure the top $k = 150$ flows. We use the ISP backbone trace
henceforth (unless mentioned otherwise),
and compare \TheSystem with the following baseline schemes.

\noindent (1) Sample and Hold: We simulate sample and hold~\cite{estan2002new}
with a flow table that is implemented as a {\em list}.  As described in
\Sec{sec:related}, we liberally allow incoming packets to look up flow entries
anywhere in the list, and add new entries up to the total memory size. The
sampling probability is set according to the available flow table size,
following recommendations from \cite{estan2002new}.

\noindent (2) Count-min sketch augmented with a `heavy flow' cache: We simulate
the count-min sketch~\cite{cormode2005improved}, but use a flow cache to keep
flow keys and exact counts starting from the packet where a flow is {\em
  estimated to be heavy} from the sketch.\footnote{A simpler
  alternative---mirroring packets with high size estimates from the count-min
  sketch---requires collecting $\approx$ 40\% of the packets in the data center
  trace played at the link rate.}
We liberally allow the count-min
sketch to use the offline-estimated exact count of the $k$th heaviest
flow in the trace, as a threshold to identify heavy flows. The flow cache is
implemented as a hash table that retains only the pre-existing flow key on a
hash collision. We allocate half the available memory each to the sketch and the
flow cache.\footnote{We follow guidance from \cite[page
    288]{estan2002new} to split the memory.}

\begin{figure}[h]
\includegraphics[max height=11cm,max width=8cm]{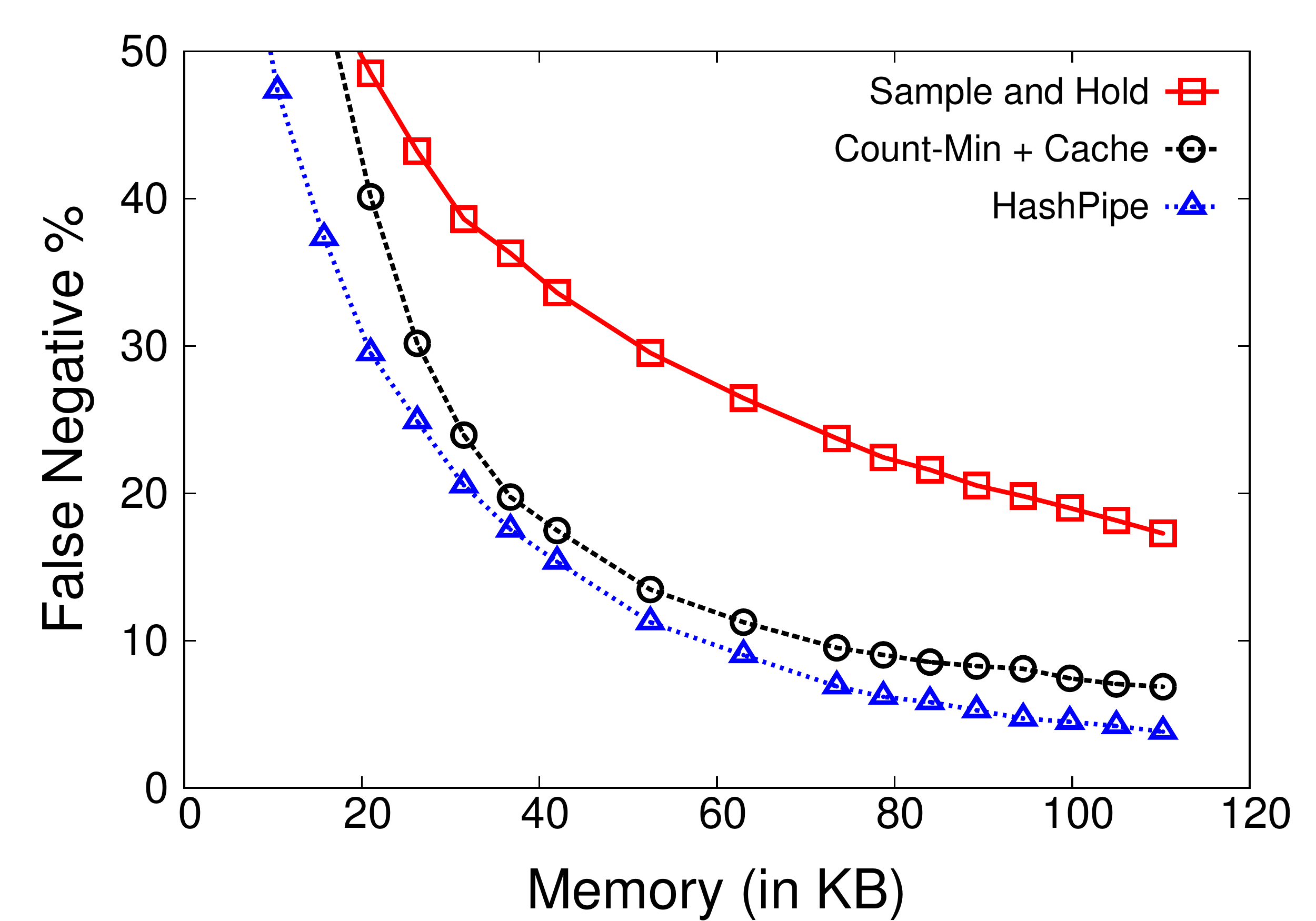}
\caption{False negatives of \TheSystem and other
  baselines. \TheSystem outperforms sample and hold and count-min sketch over
  the entire memory range.}
\label{fig:FalseNegvsMSchemes}
\end{figure}

%% In order to evaluate our scheme against our two main competitors, we ran
%% \TheSystem, Count-Min Sketch with a cache and Sample and Hold on the same traces
%% with the same amount of memory (ranging from $5KB$ to $80KB$) and compared the
%% average rate of false negatives that each of the schemes produced across all the
%% traces in detecting the top $150$ flows. In the case of \TheSystem, the memory
%% provisioned is directly translated into a certain number of entries (flowid,
%% packet count pairs) in the hash table assuming that the flow-id is the 5-tuple
%% for an Ipv4 packet. For Sample and Hold, this translates into a similar number
%% of entries based on the same principle, but this is the maximum number of
%% entries that the flow memory can hold \cite{estan2002new}. For the count-min
%% sketch combined with cache, we allocate half of the memory to the sketch itself
%% and the other half to track the heavy-hitter flows in the cache, as suggested in
%% the analysis of the Multi-stage filter \cite{estan2002new}.

\NewPara{False negatives.} \Fig{FalseNegvsMSchemes} shows false negatives
against varying memory sizes, for the three schemes compared. We see that
\TheSystem outperforms sample and hold as well as the augmented count-min sketch
over the entire memory range. (All schemes have smaller errors as the memory
increases.) Notably, at 100KB memory, \TheSystem has 15\% smaller false negative
rate than sample and hold. The count-min sketch tracks the error rate of
\TheSystem more closely from above, staying within a 3-4\% error
difference. Next, we understand where the errors occur in the baseline schemes.

\begin{figure}[h]
\includegraphics[max height=11cm,max width=8cm]{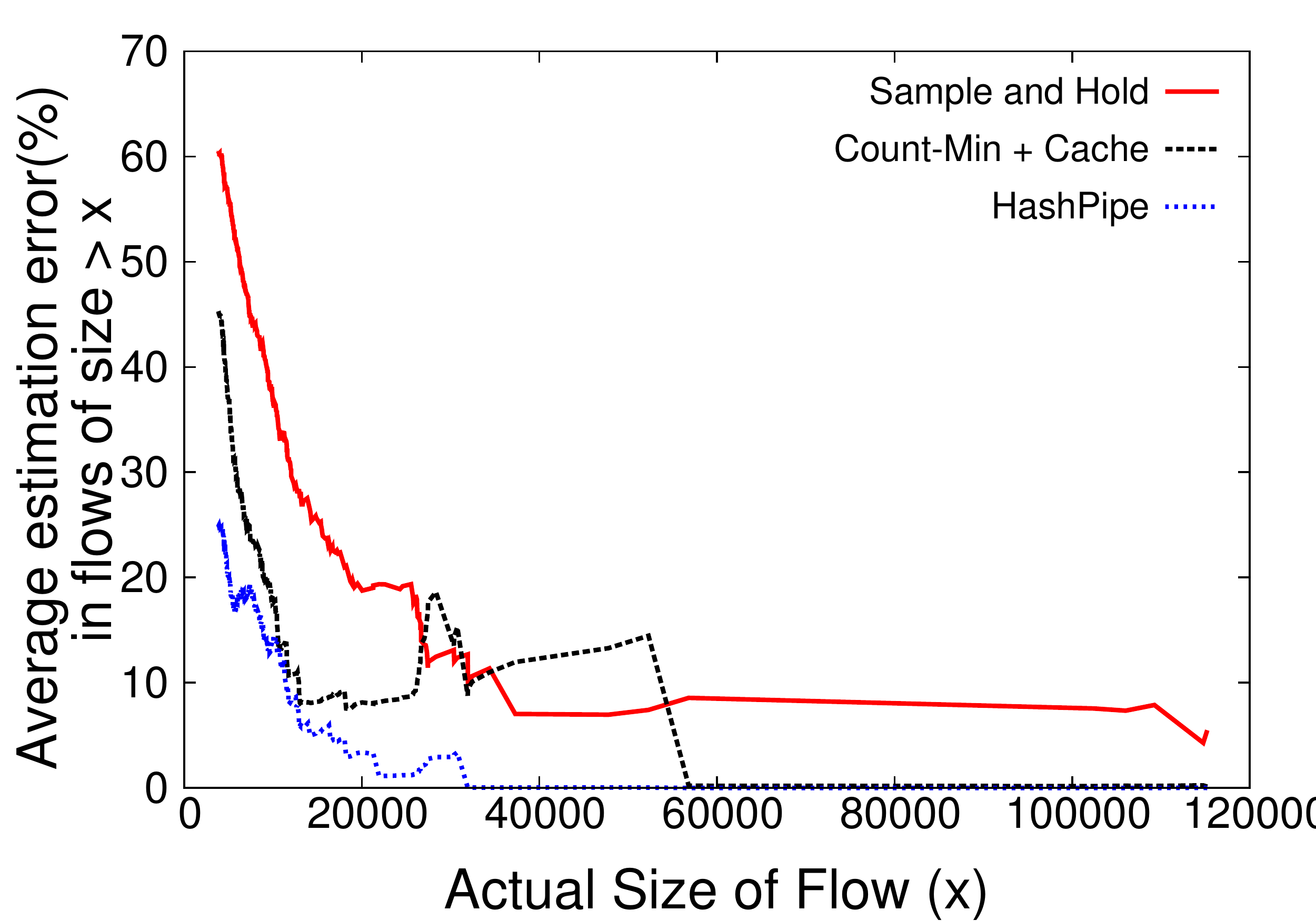}
\caption{Comparison of average estimation error (\%) of flows whose true counts
  are higher than the $x$-value in the graph. Sample and hold underestimates
  heavy flows due to sampling, and count-min sketch misses heavy flows due to
  lighter flows colliding in the flow cache. \TheSystem has no estimation errors for
  flows larger than 30,000 packets.}
  %% Comparison of average relative error in the reported sizes of flows whose
  %% actual size is larger than a given actual flow size, when detecting $150$
  %% Heavy Hitters from $0.5M$ flows on average with $26KB$ ofmemory}
\label{fig:relative-error}
\end{figure}

\NewPara{Where are the errors in the other baselines?} \Fig{relative-error}
shows the count estimation error (\%) averaged across flows whose true
  counts are higher than the $x$-value, when running all the schemes with 26KB
  memory.

Sample and hold can make two kinds of estimation errors. It can miss the first
set of packets from a heavy flow because of not sampling the flow early enough,
or (less likely) miss the flow entirely due to not sampling or the flow table
becoming full. \Fig{relative-error} shows that sample and hold makes
the former kind of error even for flows with very high true counts. For
example, there are relative errors of about 10\% even for flows of size more
than 80,000 packets. As \Fig{FalseNegvsMSchemes} shows, the errors becomes less
prominent as the memory size increases, since the sampling rate increases too.

The count-min sketch makes errors because of its inability to discriminate
between heavy and light flows during hash collisions in the sketch. This means
that a {\em light} flow colliding with heavy flows may occupy a flow cache
entry, preventing a heavier flow later on from entering the flow cache. For
instance in \Fig{relative-error}, even flows as large as 50,000 packets can have
estimation errors close to 20\%. However, as \Fig{FalseNegvsMSchemes} shows, the
effect of hash collisions becomes less significant as memory size increases.

On the other hand, \TheSystem's average error on flows larger than 20,000
packets---which is 0.2\% of the total packets in the interval---is negligible
(\Fig{relative-error}). \TheSystem has 100\% accuracy in estimating the count of
flows larger than 30,000 packets.

%% Figure \ref{fig:relative-error} analyzes the average relative error in the
%% reporting of all flows larger than a certain size $x$ across all the
%% schemes. Here, the relative error for a given flow of actual size $s_i$ and
%% reported size $r_i$ computed as $|\frac{s_i - r_i}{s_i}|$. The average
%% relative error at size $s$ is then computed as the average among the relative
%% error of all flows that are as large as and larger than $s$. For example, at
%% $x = 20K$ (which is $0.2\%$ of the total number of packets seen in this
%% interval)the average error in the reported size of all flows larger than
%% $20K$ packets is almost negligible with \TheSystem, around $10\%$ of $20K$
%% \ie $2K$ packets with the Count-Min sketch and a cache, and around $20\%$ or
%% $4K$ packets with Sample and Hold. WHle all schemes report higher flows with
%% better accuracy, \TheSystem achieves the almost $100\% $accuracy for all
%% flows of size larger than $30K$, which is faster than the other two. This
%% accounts for its ability to report the heavy hitters more accurately than
%% either of the other two.

\subsection{\TheSystem\ \vs Idealized Schemes}\label{subsec:comparisonIdeal}

We now compare \TheSystem against the idealized algorithms it is derived from
(\Sec{sec:algorithm}), namely \spacesaving~\cite{metwally2005efficient} and
\Baseline.

There are two reasons why \TheSystem may do badly relative to \spacesaving: (i)
it may evict a key whose count is much higher than the table minimum, hence
missing heavy items from the table, and (ii) it may allow too many duplicate
flow keys in the table (\Sec{sec:feed-forward}), reducing the memory available
for heavy flows, and evict heavy flows whose counts are underestimated due to
the duplicates. We showed that duplicate keys are not very prevalent in
\Sec{subsec:sensitivity}; in what follows, we understand their effects on false
negatives in the reported flows.

\begin{figure}[h]
\includegraphics[max height=11cm,max width=8cm]{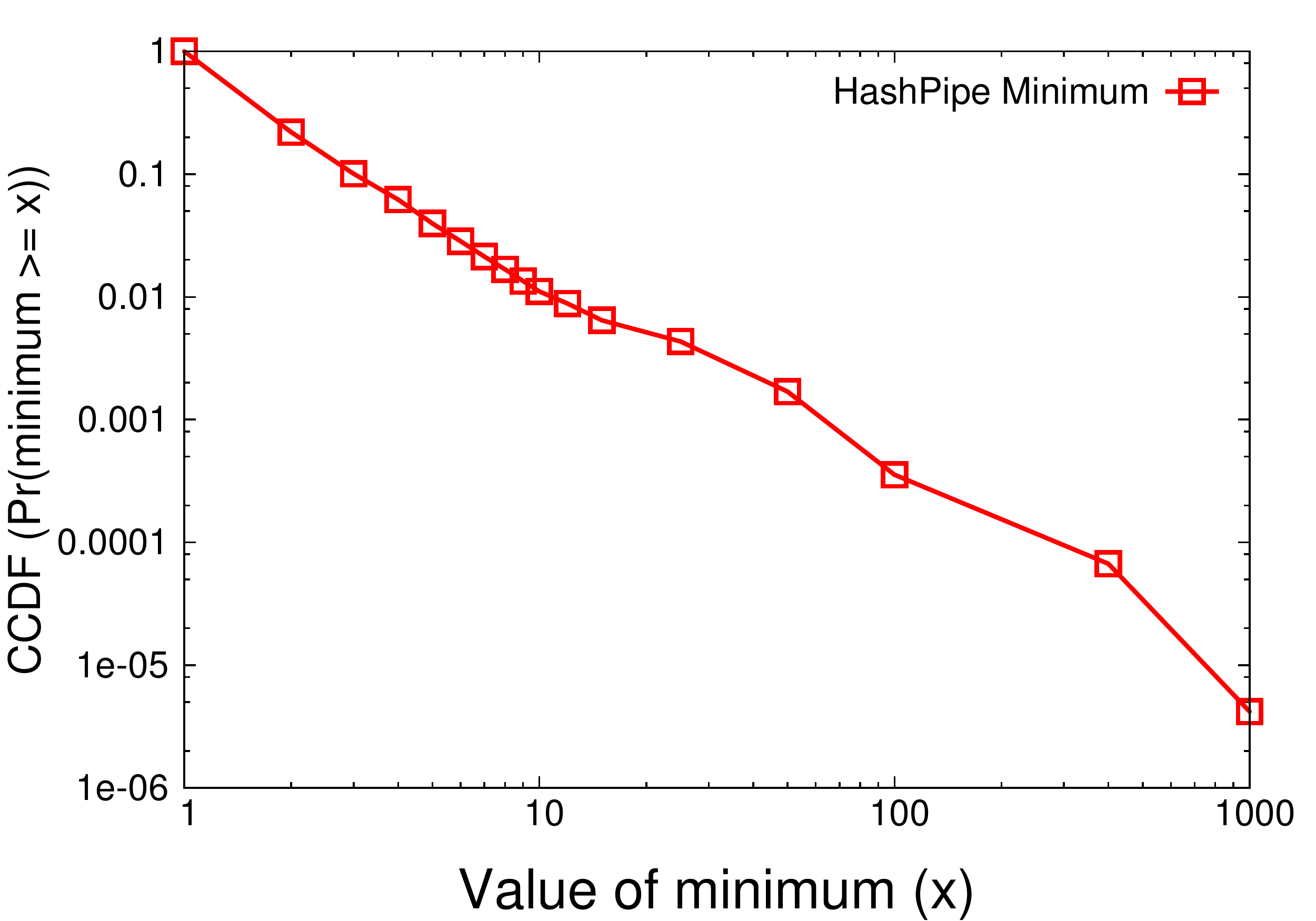}
\caption{Complementary CDF of the minimum chosen by \TheSystem on $26KB$ of
  memory, sampled every 100th packet from a trace of 10 million packets.}
\label{fig:mindist}
%plot CM, UnivMon, Reversible
\end{figure}

\NewPara{How far is the subsampled minimum from the true table minimum?}
\Fig{mindist} shows the complementary CDF of the minimum count that was chosen
by \TheSystem, obtained by sampling the algorithm's choice of minimum at every
100th packet in a 10 million packet trace. We don't show the corresponding CCDF
for the absolute minimum in the table, which only takes on two values---0 and
1---with the minimum being 1 more than 99\% likely.\footnote{Note that this is
  different from the minimum of \spacesaving, whose distribution contains much
  larger values.}
We see that the chance that \TheSystem chooses a certain minimum value decreases
rapidly as that value grows, judging from the straight-line plot on log scales
in \Fig{mindist}.
For example, the minimum counter has a value higher than 5 less than 5\% of the
time. There are a few larger minimum values (\eg 100), but they are rarer (less
than 0.01\% of the time). This suggests that it is unlikely that \TheSystem
experiences large errors due to the choice of a larger minimum from the table,
when a smaller counter exists.
%% srinivas: we haven't discussed why the absolute minimum distribution is so
%% skewed: possibly because we always insert keys with values 1. It seems less
%% important, though.

%% \TheSystem aims to model the Space Saving \cite{metwally2005efficient}
%% algorithm in a hardware-amenable version by sampling a few locations to
%% estimate the minimum instead of computing the global minimum. A natural
%% question to ask is how far are we in our minimum approximation fromt the
%% global minimum. We compared the minimum picked by \TheSystem with the actual
%% minimum in the table at the same-time for the hash table update corresponding
%% to every $100_{th}$ packet in the trace. Figure \ref{fig:mindist} shows the
%% complementary cumulative distribution function for the two minima. The two
%% distributions look reasonably similar indicating that the approximated
%% minimum isn't that far off from the actual minimum and it is very rarely
%% higher than $10$. Further, the true minimum of the table is almost always $1$
%% which is also expected since we insert flows with value $1$ in the first
%% table stage. Our scheme also reports $1$ as the actual minimum majority of
%% the time.

\begin{figure}[h]
\includegraphics[max height=11cm,max width=8cm]{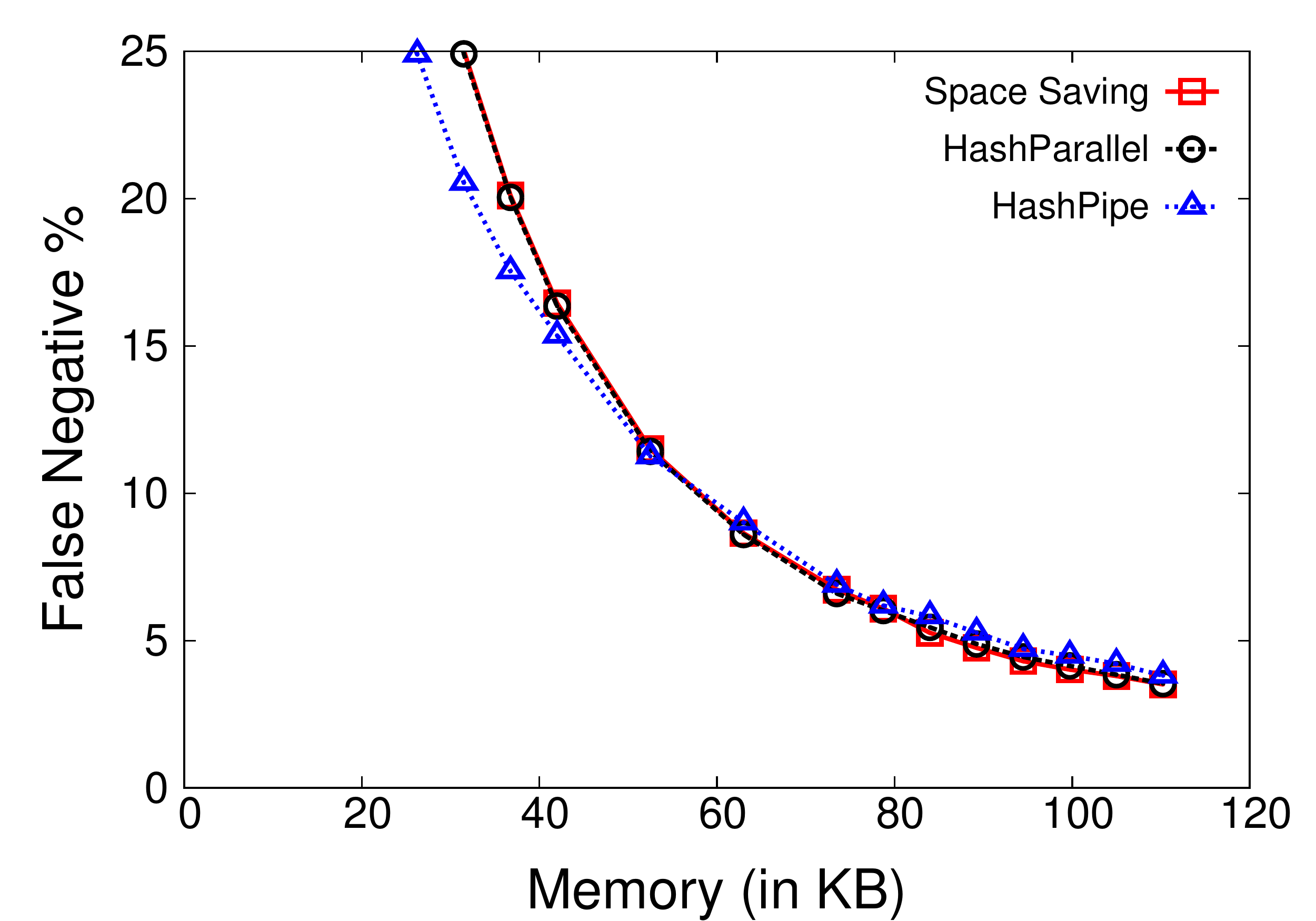}
\caption{Comparison of false negatives of \TheSystem to idealized schemes when
  detecting $k$=150 heavy hitters. In some memory regimes, \TheSystem may
  even outperform \spacesaving, while \Baseline closely tracks \spacesaving.}
%% \caption{False Negative Rate for memory sizes when detecting $150$ heavy hitters from $0.5M$ flows across $10M$ packets}
\label{fig:FalseNegvsFriendSchemes-k150}
%plot CM, UnivMon, Reversible
\end{figure}

\begin{figure}[h]
\includegraphics[max height=11cm,max width=8cm]{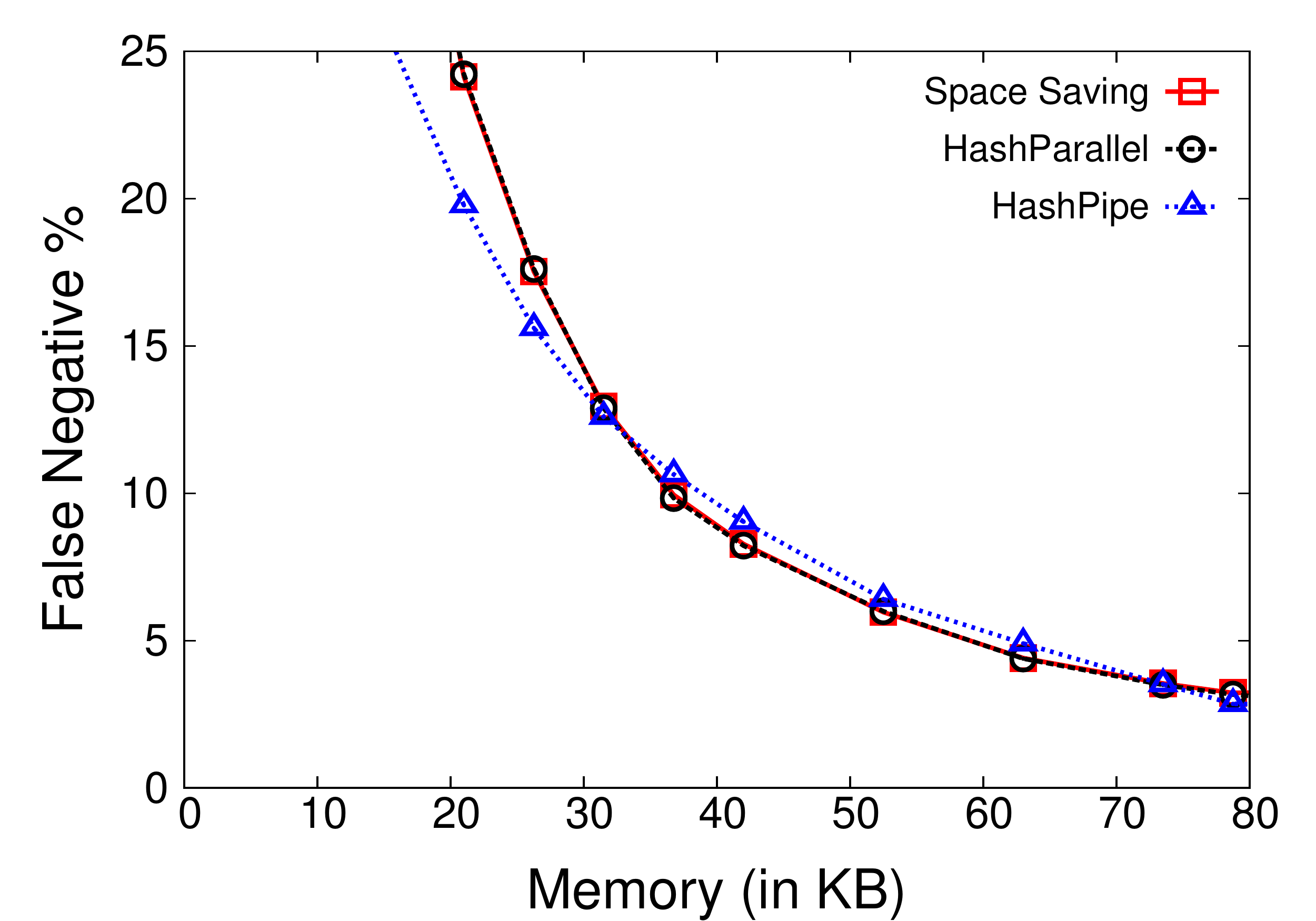}
\caption{Comparison of false negatives of \TheSystem to idealized schemes when
  detecting $k$=60 heavy hitters.}
%% \caption{False Negative Rate for memory sizes when detecting $60$ heavy hitters
%%   from $0.5M$ flows across $10M$ packets}
\label{fig:FalseNegvsFriendSchemes-k60}
%plot CM, UnivMon, Reversible
\end{figure}

\NewPara{Comparison against \spacesaving and \Baseline.} We compare the false
negatives of the idealized schemes and \TheSystem against varying memory sizes
in \Fig{FalseNegvsFriendSchemes-k150} and \Fig{FalseNegvsFriendSchemes-k60},
when reporting two different number of heavy hitters, $k$=150 and $k$=60
respectively. Two features stand out from the graph for both $k$ values. First,
wherever the schemes operate with low false negative error (say less than 20\%),
the performance of the three schemes is comparable (\ie within 2-3\% of each
other). Second, there are small values of memory where \TheSystem\ {\em
  outperforms} \spacesaving.

Why is \TheSystem outperforming \spacesaving? \Spacesaving only guarantees that
the $k$th heaviest item is in the table when the $k$th item is larger than the
average table count (\Sec{sec:spacesaving}). In our trace, the 60th item and
150th item contribute to roughly 6000 and 4000 packets out of 10 million
(resp.), which means that they require at least\footnote{10 million / 6000
  $\approx$ 1700; 10 million / 4000 $\approx$ 2500.} 1700 counters and 2500
counters in the table (resp.). These correspond to memory sizes of 30KB and 45KB
(resp.). At those values of memory, we see that \spacesaving starts
outperforming \TheSystem on false negatives.

We also show why \spacesaving fails to capture the heavier flows when it is
allocated a number of counters smaller than the minimum number of counters
mentioned above. Note that \TheSystem attributes every packet to its flow entry
correctly (but may miss some packets entirely), since it always starts a new
flow at counter value 1. However, \spacesaving increments {\em some counter} for
every incoming packet (\Alg{spacesaving}). In contexts where the number of
active flows is much larger than the number of available counters (\eg 400,000
flows with 1200 counters), this can lead to some flows having enormously large
(and grossly overestimated) counters. In contrast, \TheSystem keeps the counter
values small for small flows by evicting the flows (and counts) entirely from
the table.

\begin{figure}[h]
\includegraphics[max height=11cm,max width=8cm]{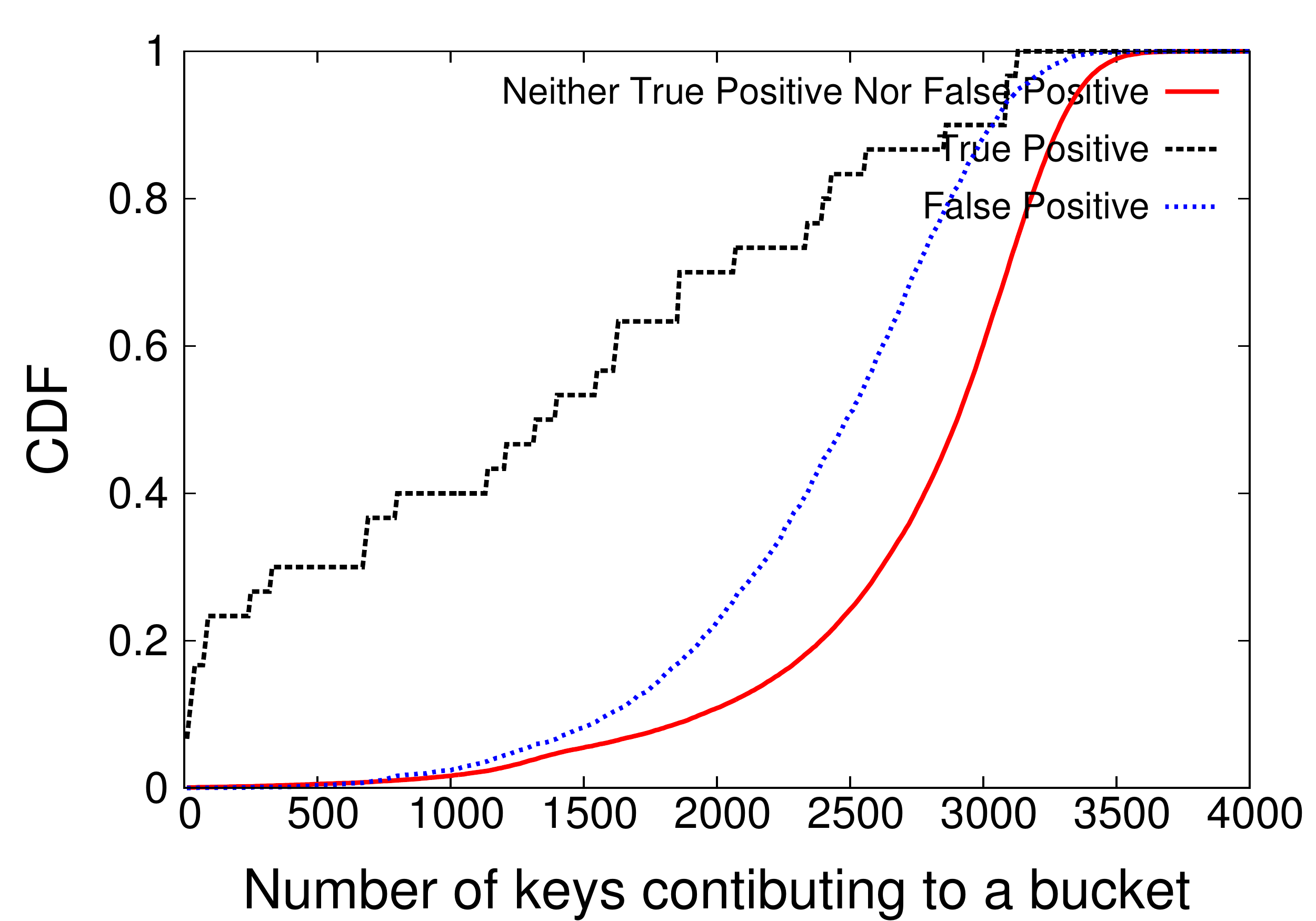}
\caption{CDF of the number of distinct flows contributing to a counter in
  \spacesaving's table, when identifying $k$=150 heavy flows with $m$=1200
  counters. We show three distributions according to the flow's label after the
  experiment.}
%% \caption{Chances of a given bucket having a certain number of keys contributing
%%   to it when running \spacesaving to identify $150$ heavy hitters with $1200$
%%   counters. Legend represents category that the flow found in a bucket at the
%%   end of the interval belongs to.}
\label{fig:SSkeysperbucket}
\end{figure}

At memory sizes smaller than the thresholds mentioned above, incrementing a
counter for each packet may result in {\em several small flows} catching up to a
heavy flow, leading to significant false positives, and higher likelihood of
evicting truly heavy flows from the table. We show this effect on \spacesaving
in \Fig{SSkeysperbucket} for $k$=150 and $m$=1200 counters, where 
in fact $m=2500$ counters are required as described earlier.
The distribution of the number of keys contributing to a false positive flow
counter in the table is clearly shifted to the right relative to the
corresponding distribution for a true positive.
%
%% A similar and
%% natural characterization of the distributions holds for the estimation errors
%% from the different categories of flow keys, shown in \Fig{SSdevperbucket}.

%% To quantify how we do compared to \spacesaving, we compared the performance of
%% \TheSystem against the intermediate \Baseline as well as \spacesaving in terms
%% of the false negative rate in the reporting of $150$ heavy hitters with
%% different amounts of memory. As reported in Figure
%% \ref{fig:FalseNegvsFriendSchemes}, \Baseline is almost identical to \spacesaving
%% in its performance. This is not surprising since Figure \ref{fig:mindist} shows
%% that sampling the minimum doesn't result in such a different distribution from
%% the actual minimum. However, \TheSystem has significantly lower false negative
%% rates (by almost $15\%$) than either of them at almost all memory sizes. This is
%% attributed to the fact that \TheSystem inserts a new flow with value $1$ in the
%% first table while \spacesaving and \Baseline both replace it with the current
%% minimum value incremented by $1$.

\begin{figure}[h]
\includegraphics[max height=11cm,max width=8cm]{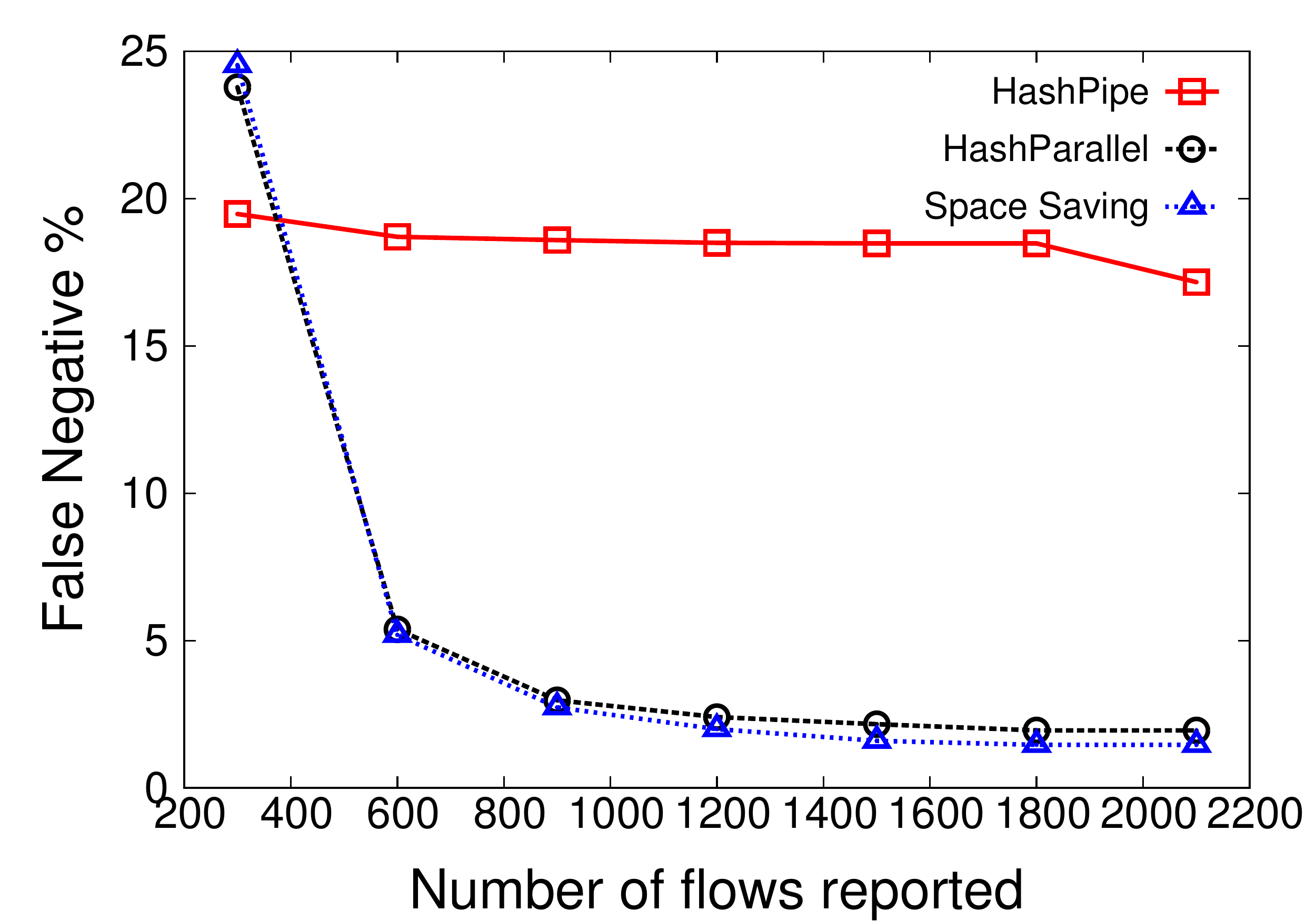}
\caption{Benefits of overreporting flows on false negative errors
  with $k$=300 heavy flows and $m$=2400 counters. While \spacesaving and
  \Baseline improve significantly by reporting even $2k$ flows,
  \TheSystem does not, because of evictions due to duplicate entries.}
%% \caption{
%%   Comparison of false negatives of HashPipe to \spacesaving and Hash
%%   Parallel when reporting more than $k$ candidate flows. \spacesaving and
%%   HashParallel significantly outperform HashPipe even with $2k$ reported flows}
\label{fig:reportingMoreThanK-falseneg}
\end{figure}

\NewPara{Impact of duplicate keys in the table.} Finally, we investigate how
duplicate keys and consequent underestimation of flow counts in the table may
affect the errors of \TheSystem. In \Fig{reportingMoreThanK-falseneg}, we show
the benefits of reporting {\em more than $k$ counters} on false negatives, when
the top $k$=300 flows are requested with a memory size of $m$=2400 counters.
While the false negative rates of \spacesaving and \Baseline improve
significantly with overreporting, the errors for \TheSystem remains flat
throughout the interval, dropping only around 1800 reported flows. We infer that
most heavy flows are retained somewhere in the table for \spacesaving and
\Baseline, while \TheSystem underestimates keys sufficiently often that they are
completely evicted---to the point where overreporting does not lower the false
negative errors.
%
%% \Fig{reportingMoreThanK-falsepos} shows that 
We find that overreporting
flows only increases the false positive errors slightly for all schemes, with
values staying between 0.1-0.5\% throughout.

\iffalse
\begin{figure}[h]
%width=.49\columnwidth
\includegraphics[width=0.95\columnwidth,max height=11cm,max width=8cm]{FriendsFalsePosWithRep.pdf}
\caption{Impact of overreporting flows from the table on false positive errors
  with $k$=300 heavy flows and $m$=2400 flows. %%The errors stay within a small
  %%range.}
  }
\label{fig:reportingMoreThanK-falsepos}
\end{figure}
\fi

%%
%- scatter plots to show where errors arise in our scheme. Sources of error include effect of hash collisions and packet interleaving. Possibly talk about prevalence of duplicates in sequential probing as a source of error. (generate graph here)
%- impact of (i) subsampling the table (as opposed to looking at the entire table), assuming you see all $D$ probes at once (ii) partial knowledge/sequential lookup instead of seeing all $D$ locations at once.
%(the two categorizations above may lead to the same!)
%%

%%

%zoomed in only top 1000 flows
%Focusing on only the top 1000 flows
%\includegraphics[max height=11cm,max width=8cm]{image031}
%\includegraphics[max height=11cm,max width=8cm]{image033}

%% file: other-related.tex
\section{Related Work}
\label{sec:other-related}

\NewPara{Applications that use heavy hitters.} Several applications
use information about heavy flows to do better traffic
management or monitoring. DevoFlow~\cite{devoflow} and
Planck~\cite{planck} propose exposing heavy flows with low overhead as ways to
provide better visibility and reduce congestion in the
network. UnivMon~\cite{univmon} uses a top-$k$ detection sketch internally as a
subroutine in its ``universal'' sketch, to determine more general statistics
about the network traffic. There is even a P4 tutorial application on switch
programming, that performs heavy hitter detection using a count-min-sketch-like
algorithm~\cite{p4-sigcomm-tutorial-hh-example}.

\NewPara{Measuring per-flow counters.} Prior works such as
FlowRadar~\cite{li2016flowradar} and CounterBraids~\cite{counterbraids}
%% others, \eg~\cite{li-randomized-counter-sharing}
have proposed schemes to
measure accurate per-flow traffic counters. Along similar lines, hashing schemes
like cuckoo hashing~\cite{cuckoo-hashing} and d-left
hashing~\cite{vocking2003asymmetry} can keep per-flow state memory-efficiently,
while providing fast lookups on the state. Our goal is not to measure or keep
all flows; just the heavy ones. We show (\Sec{sec:evaluation}) that \TheSystem
uses 2-3 orders of magnitude smaller memory relative to having per-flow state
for all active flows, while catching more than 90\% of the heavy flows.

\NewPara{Other heavy-hitter detection approaches.} The multi-resolution tiling
technique in ProgME~\cite{progme}, and the hierarchical heavy-hitter algorithm
of Jose \etal~\cite{jose2011online} solve a similar problem as
ours. They estimate heavy hitters in traffic {\em online}, by iteratively
``zooming in'' to the
portions of traffic which are more likely to contain heavy flows. However, these
algorithms involve the control plane in running their
flow-space-partitioning algorithms, while \TheSystem works completely within the
switch pipeline. Further, both prior approaches require temporal stability of
the heavy-hitter flows to detect them over multiple intervals; \TheSystem
determines heavy hitters using counters maintained in the same interval.

%% file: conclusion.tex
\section{Conclusion}\label{sec:conclusion}

%In the paper, we studied the constraints switch architectures apply on
%algorithm implementations in order to maintain high packet processing.
%Focusing on heavy hitter detection, we design an algorithm that follows these
%restrictions while still achieving high accuracy.

In this paper, we proposed an algorithm to detect heavy traffic flows within the
constraints of emerging programmable switches, and making this information
available within the switch itself, as packets are processed. Our solution,
\TheSystem, uses a pipeline of hash tables to track heavy flows preferentially,
by evicting lighter flows from switch memory over time. We prototype \TheSystem
with P4, walking through the switch programming features used to implement our
algorithm. Through simulations on a real traffic trace, we showed that
\TheSystem achieves high accuracy in finding heavy flows within the memory
constraints of switches today.%%  while outperforming existing heavy-hitter
%% detection schemes that use sampling and sketching.
%% In this paper, we discussed the constraints imposed by switching architectures on
%% algorithmic implemenations due to the need for high speed packet processing. In
%% particular, we propose an algorithm for heavy hitter detection called \TheSystem
%% that respects these restrictions while still achieving high accuracy. \TheSystem
%% uses a pipeline of hash tables along with a local minimum computation at each
%% pipeline stage to preferentially track the heavier flows. We also prototype the
%% algorithm in P4. Through our simulations, we show that \TheSystem achieves lower
%% error rates than existing sampling and sketching based approaches %like Sample
%%                                 %and Hold and Count-Min Sketch 
%% at comparable amounts of memory.
%% We are further studying \TheSystem through synthetic workloads, additional
%% optimizations to the algorithm, and theoretical analyses.

%univmon, dataplane
% assymmetric memory split
%synthetic workloads
%compiling to a target
%extensive analysis